\documentclass[journal]{IEEEtran}

\usepackage{amsmath,amssymb,amsfonts}
\usepackage{algorithmic}
\usepackage{graphicx}
\usepackage{float}
\usepackage{textcomp}
\usepackage{wrapfig,colortbl}
\usepackage[font=small]{caption, subcaption}
\usepackage{multirow}
\usepackage{booktabs}
\usepackage{multicol}

\begin{document}

\title{Optically-Validated Microvascular Phantom for\\ Super-Resolution Ultrasound Imaging}

\author{Jaime Parra Raad$^1$, Daniel Lock$^1$, Yi-Yi Liu$^2$, Mark Solomon$^2$, Laura Peralta$^3$, and Kirsten Christensen-Jeffries$^1$
\thanks{This research was supported by the Medical Research Council (MR/S023542/1).}
\thanks{L. Peralta acknowledges financial support from the Royal Society (URF/R1/211049).}
\thanks{$^1$ Research Department of Imaging Physics and Sciences, King's College London, London, United Kingdom}
\thanks{$^2$ Department of Biomedical Engineering and Imaging Sciences, King's College London, London, United Kingdom}
\thanks{$^3$ Department of Surgical and Intervention Engineering, King's College London, London, United Kingdom}}

\maketitle

\begin{abstract}
Super-resolution ultrasound (SRUS) visualises microvasculature beyond the ultrasound diffraction limit (wavelength($\mathbf{\lambda}$)/$\mathbf{2}$) by localising and tracking spatially isolated microbubble contrast agents. SRUS phantoms typically consist of simple tube structures, where diameter channels below $\mathbf{100}$ $\mathbf{\mu}$m are not available. Furthermore, these phantoms are generally fragile and unstable, have limited ground truth validation, and their simple structure limits the evaluation of SRUS algorithms. To aid SRUS development, robust and durable phantoms with known and physiologically relevant microvasculature are needed for repeatable SRUS testing. This work proposes a method to fabricate durable microvascular phantoms that allow optical gauging for SRUS validation. The methodology used a microvasculature negative print embedded in a Polydimethylsiloxane to fabricate a microvascular phantom. Branching microvascular phantoms with variable microvascular density were demonstrated with optically validated vessel diameters down to $\mathbf{\sim60}$ $\mathbf{\mu}$m ($\mathbf{\lambda/5.8}$; $\mathbf{\lambda=\sim350}$ $\mathbf{\mu}$m). SRUS imaging was performed and validated with optical measurements. The average SRUS error was $\mathbf{15.61}$ $\mathbf{\mu}$m ($\mathbf{\lambda/22}$) with a standard deviation error of $\mathbf{11.44}$ $\mathbf{\mu}$m. The average error decreased to $\mathbf{7.93}$ $\mathbf{\mu}$m ($\mathbf{\lambda/44}$) once the number of localised microbubbles surpassed $\mathbf{1000}$ per estimated diameter. In addition, the less than $\mathbf{10\%}$ variance of acoustic and optical properties and the mechanical toughness of the phantoms measured a year after fabrication demonstrated their long-term durability. This work presents a method to fabricate durable and optically validated complex microvascular phantoms which can be used to quantify SRUS performance and facilitate its further development.

\textbf{Microvascular Phantom}

\vspace{0.25pc}\hspace{0pc}\includegraphics[width=3.1in]{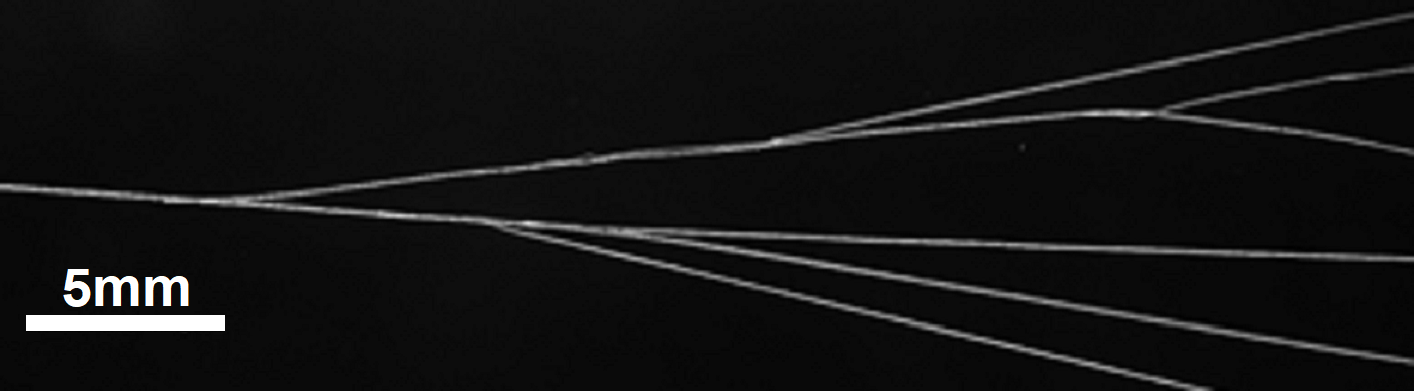}

\vspace{0.20pc}
\textbf{Super-Resolution Ultrasound Image}

\vspace{0.15pc}\hspace{0pc}\includegraphics[width=3.1in]{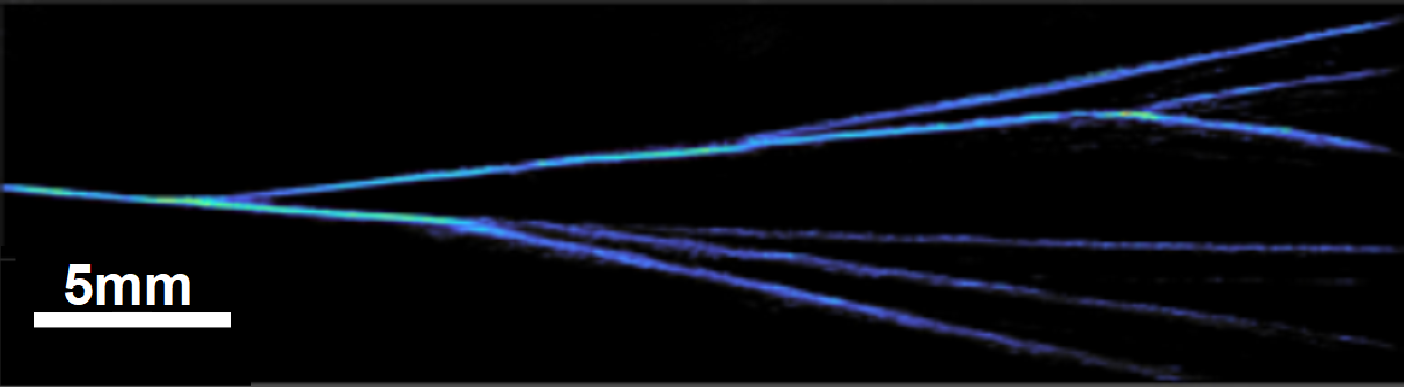}

\end{abstract}

\begin{IEEEkeywords}
Super-Resolution Ultrasound Imaging, Ultrasound Localisation Microscopy, Contrast Enhanced Ultrasound Imaging, Ultrasound Phantom, Microvascular Phantom, Vascular Phantom, Tissue-Mimicking Material, Microfluid chip.
\end{IEEEkeywords}
\bigskip
\section{Introduction}
\label{sec:introduction}

\IEEEPARstart{I}{n} the last decade, novel ultrasound (US) imaging techniques that visualise microvascular structures beyond the US diffraction limit have been developed; they are known as Super-Resolution Ultrasound (SRUS) imaging \cite{ChristensenJeffriesSRUSimaging}. The most well-known SRUS method maps these structures by localising and tracking microbubbles (MBs) flowing in the streams of its vessels, also known as US localisation microscopy \cite{ChristensenJeffriesSRUSimaging}. It has been demonstrated that SRUS is a powerful tool to describe the microvascular morphology of a mass \cite{BROWNSRUSRatHepatocellularCarcinoma, CoutureSRUSStateOfArt, SONGSRUSReadyForClinic}. Its potential to map complex microvascular structures has been shown in various in vivo models \cite{HuangSRUSChickenEnmbryo, ChristensenJeffriesSRUSimagingEarMouse, HeilesSRUSPALABrainMouse, DemeulenaereSRUSMouseBrain, ZHANGSRUSRabbitTumour}, and in human pilot studies \cite{DenisSRUSRatKidneyHumanKidney, U-WaiSRUSHumanKidney, OrSRUSHumanBreast}.

Validating SRUS results enables its improvement and is paramount to translating the technique to the clinic. Phantoms of different complexities and with geometry parts smaller than the US diffraction limit (wavelength($\lambda$)/$2$) have been used to support detailed investigations of SRUS within controlled environments. For example, single microtube phantoms have been used to study the performance of SRUS algorithms under different conditions. Christensen-Jeffries \textit{et al.} \cite{ChristensenSRUS3DParaffinPhantom} used a single spiral $0.2$ mm ($\sim\lambda/6$) inner diameter tube submerged in paraffin gel to investigate the performance of SRUS in 3D. O'Reilly \textit{et al.} \cite{OReillySkullPhantomSRUS} used a single spiral $0.25$ mm ($\sim\lambda/10$) inner diameter tube in an ex vivo human skull submerged in water to study the aberrating effects of the skull bone in SRUS results. Desailly \textit{et al.} \cite{DesaillySRUSLocalisationErrorPDMS} used a squared cross-section, $60$x$80$ $\mu$m$^2$ ($\sim\lambda/10$), microchannel (MC) embedded in polydimethylsiloxane (PDMS) to investigate the localisation error of SRUS. Similarly, Christensen-Jeffries \textit{et al.} \cite{ChristensenJeffriesSRUSXshapePhantom} used a simple X-shape microvascular structure comprised of two $0.2$ mm ($\sim\lambda/3.2$) inner diameter tubes submerged in water to study SRUS localisation performance. Harput \textit{et al.} \cite{HarputSRUS3DPhantom} demonstrated 3D SRUS by imaging two touching $0.2$ mm ($\sim\lambda/2$) inner diameter tubes, which were twisted together and submerged in water. The phantom proposed by Harput \textit{et al.} had a geometry with sizes smaller than the US diffraction limit in the region where the twisted tubes were touching. However, these phantoms have a simplistic structure and do not capture the flow behaviour and the spatial distribution of complex microvasculature.

In an effort to obtain more advanced microvascular structures, the use of 3D printing technology, microfluidic chips (soft lithography), or wire-pulling technology has been proposed for the fabrication of complex SRUS phantoms.  However, these methods suffer from several technical hurdles in replicating physiologically relevant MB flow and vessel distribution; the obtained microvascular structures are composed of single MCs or MCs forking, and their cross-sections are typically rectangular or a cluster of circles with minimum diameters between $\sim0.1$ to $\sim0.2$ mm. Phantoms with MC diameters of $0.2$ mm limit the SRUS performance evaluation to frequencies below $3$ MHz as their dimension is about the diffraction limit (assuming wave speed of $1500$ m/s). In addition, some use hydrogel-based tissue-mimicking materials (TMMs), which are known to have poor durability (on the order of weeks) and rigidity when stored and used at room temperature \cite{LamouchePhantomReview, CortelaHydrogelPhantomDurability, DemitriHyydrogelDurabilityPhantom}. Ommen \textit{et al.} \cite{OmmenSRUS3DPhantom}, for example, used a 3D printer to fabricate point scatterers with dimensions of $0.2$x$0.2$x$0.2$ mm$^3$ ($\sim\lambda/2.5$) embedded in hydrogel to investigate the 3D SRUS performance. Desailly \textit{et al.} \cite{DesaillyHoneycomb2DphantomSRUS} proposed a microfluidic chip comprised of MCs with a square cross-section of $40$x$80$ and $100$x$80$ $\mu$m$^2$ ($\sim\lambda/20$ and $\sim\lambda/10$) forming a "honeycomb" structure to investigate the performance of SRUS. The microfluidic chip was printed using soft lithography in PDMS. Shangguan \textit{et al.} \cite{ShangguanForkedSRUSPhantom} proposed the use of additive manufacturing of Inconel 625 and Polyvnyl-Alcohol-based to form a SRUS phantom with forked shape MCs. The MCs cross-section was a rectangular shape of $2$x$0.25$ mm$^2$ ($\sim\lambda/0.63$). Their forked junction provided an area with geometry dimensions smaller than their US diffraction limit. They evaluated SRUS performance on the MCs and obtained an error of 68 $\mu$m. Kawara \textit{et al.} \cite{KawaraBifurcationPhantom} proposed pulling adjacent wires in a hydrogel material to form a microvascular phantom. The vessel cross-sections were shaped according to the number of adjacent wires used to make them. SRUS results were demonstrated on a single MC of $0.1$ mm in diameter (imaging frequency $7.8$ MHz, material wave speed not reported). None of the phantoms previously described presented validation measurements of the MC dimensions reported aside from the one presented by Kawara et al. \cite{KawaraBifurcationPhantom}.

In vivo models comprise realistic microvascular flow and MB dynamics. However, their validation is challenging due to movement, microvascular changes in time, and the lack of in vivo imaging techniques that visualise microvasculature to the micrometre. To validate SRUS in vivo, qualitative comparisons with a second imaging modality have been investigated, e.g. comparing to optical images \cite{ChristensenJeffriesSRUSimagingEarMouse} or in vivo micrometre computational tomography (micro-CT) \cite{BROWNSRUSRatHepatocellularCarcinoma}. Overall image quantitative assessment using ex-vivo micro-CT \cite{AndersenSRUSRatKidneyCT,ChenSRUSCTValidationExvivo} has demonstrated anatomical validation of the SRUS results. However, in some cases, comparing the same microvessel using both imaging modalities was challenging due to tissue deformation and the 3D nature of the microvasculature. 

The existing strategies proposed to validate SRUS performance either mimic flow velocities and acoustic tissue characteristics in vitro without vascular complexities, such as vascular branching or variable microvascular density, or they do not allow rigorous quantitative SRUS validation in vivo. In addition, the phantoms and the in vivo models used are not long-lasting; they cannot be used over time to track improvements of new US data acquisition methods and SRUS strategies. A durable system that can be used repetitively, mimics in vivo microvasculature and allows the quantification of SRUS performance would help translate SRUS imaging to the clinic.

This paper presents a reproducible methodology to fabricate durable and optically validated microvascular phantoms and demonstrates their use to quantitatively measure SRUS performance. The methodology allows the fabrication of microvascular structures with vascular branching and variable vessel density across the phantom. The fabricated phantoms contain a hollow and wall-less microstructure composed of a single vessel branching into two or more vessels of different diameters, with a minimum diameter of $\sim\lambda/5.8$. The microvascular structure in the phantom was inspired by branching in vivo microvasculature \cite{ChristensenJeffriesSRUSimagingEarMouse}, where the main vessel bifurcates into two secondary vessels, and each of the secondary vessels branches into three tertiary vessels. The vessels that form the microstructure are of circular cross-section, mimicking in vivo blood vessels. PDMS was used as TMM to fabricate the microvascular phantom. In addition, a methodology to evaluate SRUS performance based on the number of localisations per estimated diameter is presented. The performance evaluation methodology is based on the trueness and precision defined by ISO \cite{ISO5725-1}.

PDMS was used as a TMM due to its easy manufacturing process, excellent optical transparency, and durability, which allows the fabrication of complex microfluidic chips \cite{CHENPhantomMaterialsAgarPDMS,WangPhantomPDMSManufacturing, FeltonPhantomPDMSMicrofluidic}. PDMS has been used in the past for the fabrication of MC structures. Wang \textit{et al.} \cite{WangPhantomPDMSManufacturing} and Felton \textit{et al.} \cite{FeltonPhantomPDMSMicrofluidic} proposed the use of dissolvable 3D printed parts and PDMS to form microfluidic chips. They obtained MC with a square cross-section of $0.2$x$0.2$ mm$^2$. In addition, PDMS has been used to fabricate vascular phantoms in the past. Long \textit{et al.} \cite{LongPAPhantomPDMS} used soft lithography to obtain a porcine liver optical phantom with a minimum MC diameter of $\sim0.1$ mm. Wu \textit{et al.} \cite{WuMicrovascularPhantomPDMS} proposed the use of soft lithography on PDMS to fabricate tumour hypoxia and vascular anomalies in photoacoustic phantoms. 

The following sections present the microvascular phantom fabrication method, the SRUS imaging procedure, the resulting phantoms, the SRUS results, and a discussion of the outcomes. 
\section{Microvascular Phantom Fabrication Methodology}
\label{sec:method}
\subsection{Tissue Mimicking Material Characterisation}
\label{sec:tmm}

PDMS (PDMS, CAS: 100-41-4, Dow Sylgard, UK) TMM matrix was fabricated following the formulation suggested by the seller. The base and the curing agent were mixed in a $10:1$ ratio at room temperature. The mixture was degassed in a vacuum chamber for $5$ min and poured into a mould. The mix in the mould was degassed again to remove any air bubbles created in the pouring process. Finally, the mixture was cured at $65^\circ C$ in an oven for $2$ hours. 

The longitudinal wave velocity, acoustic amplitude attenuation, optical opacity, and durability of PDMS were measured to evaluate its feasibility as TMM for microvascular phantoms. The material properties of PDMS and its durability were evaluated on the phantoms. Pulse-echo tests were used to estimate the wave velocity and acoustic attenuation of the TMM. The pulse-echo test consisted of a $3$ cycle, sine-weighted, $3$ MHz sinusoidal signal, excited at the sample's surface, that travelled through the material and reflected at the back wall of the sample. The US signal was excited using an Esaote LA332 US probe connected to a ULA-OP 256 system \cite{BoniULAOP256System}. The reverberations from the back wall were recorded with the ULA-OP 256 system and analysed in MATLAB \cite{MATLAB}. The wave speed was calculated as shown in equation \ref{eq:wavespeed}.

\begin{equation}
vel=\frac{2*T}{\Delta t}
\label{eq:wavespeed}
\end{equation}

Where $vel$ is the longitudinal wave velocity in PDMS, $T$ is the thickness of the PDMS sample, and $\Delta t$ is the time between two consecutive reverberations. The amplitude attenuation was estimated using the following equation:

\begin{equation}
a= \frac{P_{2} - P_1}{D_{2} - D_1}
\label{eq:attenuation}
\end{equation}

Where $a$ is the longitudinal wave attenuation in PDMS, $P$ is the peak amplitude of the reverberations envelope in decibels, $D$ is the travelled distance at which the reverberation was recorded, and the subscript number indicates reverberation recorded.

The optical opacity of the materials was estimated as the percentage of light blocked when travelling through a TMM disc sample. Two measurements were performed per TMM to obtain the optical opacity percentage. First, the optical power of the light source was recorded using an optical power meter (PM100D, Thorlabs, Inc.). Then, the optical power after travelling through the TMM was recorded. The light source was kept constant and at the same distance from the optical power meter in both measurements. The optical power was recorded at a wavelength of $532$ nm. The optical opacity percentage was calculated as shown in equation \ref{eq:opticalopacity}. 

\begin{equation}
OP=100 \% * \left( 1-\frac{I_d}{I_s} \right)
\label{eq:opticalopacity}
\end{equation}

where $OP$ is the optical opacity, $I_s$ is the source intensity, and $I_d$ is the intensity measured after the light travels through the TMM sample. 

\subsection{Phantom Fabrication}
\label{sec:phantom_fabrication_optical_validation}

\begin{figure}[h]
\centering

        \includegraphics[width=0.9\linewidth]{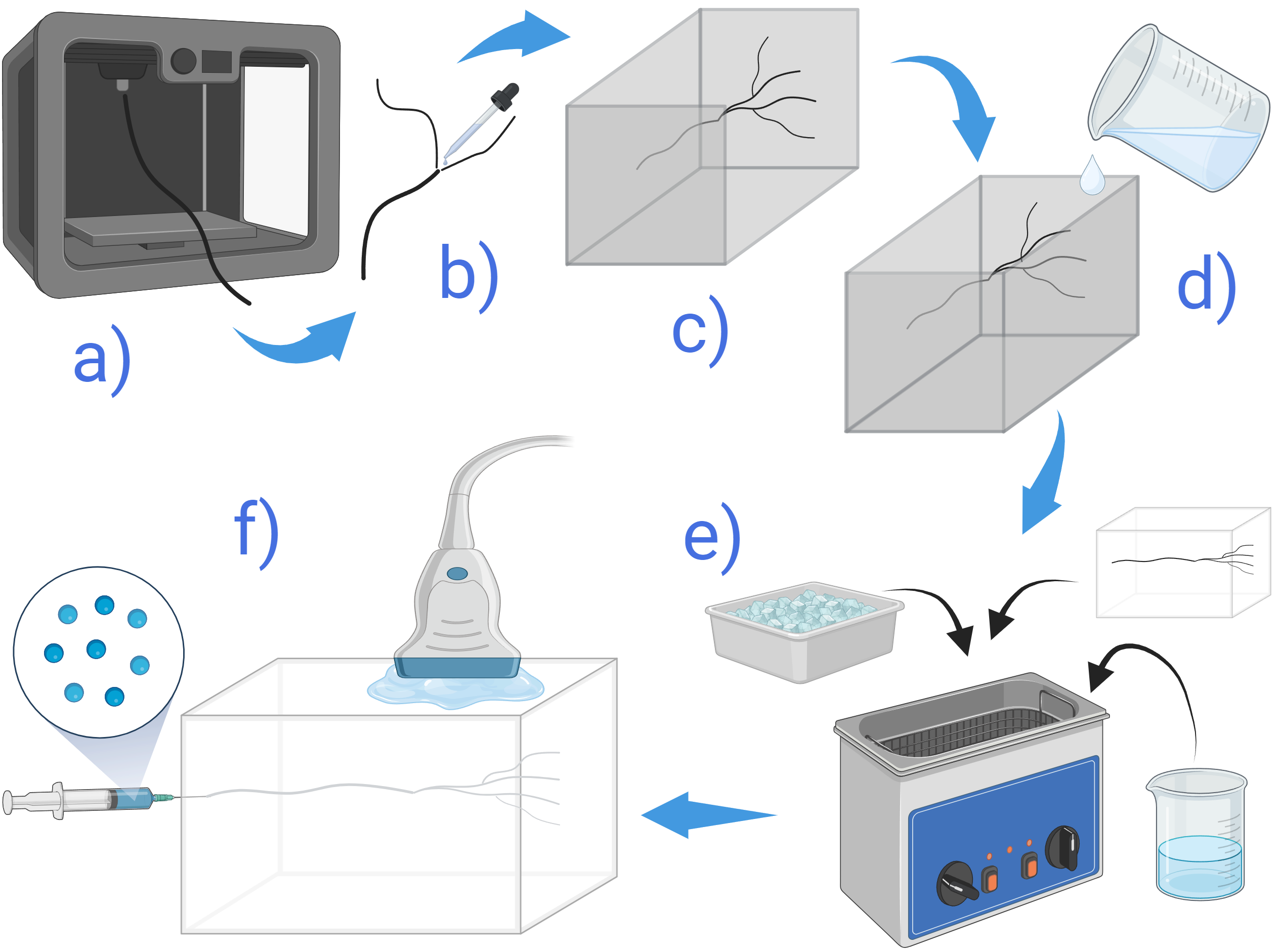}
   
\caption{Microvascular phantom manufacturing process. a) Microrods are formed by extruding thermoplastic materials using a 3D printer. b) Microrods are connected to form a negative print of the microvasculature. The microrods are connected using the solvent that dissolves the thermoplastic material. The formed negative print is placed in the phantom laser-cut mould. c) The TMM matrix is poured into the microvascular phantom mould and left to cure. e) The phantom is placed in a US cleaner to dissolve the negative print of the microvasculature. The US cleaner is filled with the solvent of the thermoplastic material used to form the negative print. Ice bags are added to the solvent to extend the cleaning period. f) MB are injected into the microvascular phantom. The MB flow is imaged with US to test the usability of the phantom. The image was created with BioRender.com.}
\label{fig:Methodology}
\end{figure}

Figure \ref{fig:Methodology} illustrates the proposed methodology to manufacture microvascular phantoms. The microrods were fabricated by extruding a thermoplastic material, Acrylonitrile Butadiene Styrene (ABS), using an Ultimaker 2+ connect 3D printer (UltiMaker, Netherlands) with a $0.6$ mm nozzle (figure \ref{fig:Methodology}a). The 3D printer was programmed to extrude the ABS continuously. Then, the extruded material was manually pulled to form microrods with diameters smaller than $0.6$ mm, ranging from $\sim50$ $\mu$m to $\sim 600$ $\mu$m. Once the microrods were formed, their diameter was measured with an optical microscope (BX3M, Evident Corp., Japan) and sorted based on size. 

Two microrods are connected using an ABS solvent, acetone (acetone, 99.5\% pure, CAS: 67-64-1, Fisher BioReagents, UK). The solvent is spread on the local surface of the connection, partially dissolving the microrods at the joint area and creating a bifurcation once the solvent is dried. Several bifurcating microrods can be connected to form a branch. Figure \ref{fig:Methodology}b illustrates the connection of two microrods using a solvent.

Several microrods were connected to form a negative print of a microvasculature. Microrods with $\sim600$ $\mu$m in diameter were then attached to the ends of the negative print. These microrods served as inlet and outlet insertions for the microvasculature. The microstructure was positioned in the phantom mould as shown in figure \ref{fig:Methodology}c. The mould was laser-cut to obtain precise alignment of the microstructure. The inlets and outlet insertions were placed in laser-cut holes in the mould, which facilitated the precise alignment of the microvascular structure. Then, the TMM matrix was poured into the mould and cured (figure \ref{fig:Methodology}d). The TMM was mixed and degassed as described in section \ref{sec:tmm}.

After the TMM was cured and supported the microvasculature negative print, the phantom was placed at the centre of a US cleaner (ultrasonic cleaner, 100W, 3L, RS Components Ltd, UK) with acetone and ice bags to dissolve the microrods (figure \ref{fig:Methodology}e). The ice allows continuous usage of the US cleaner by maintaining its temperature at $\sim25$ $^\circ$C or below. The acetone was refilled every $30$ min or whenever its level fell below the one recommended by the US cleaner manufacturer. The rectified US wave program of the US cleaner was selected to avoid forming cracks in the phantom; cracks could be formed due to the continued expansion of the TMM. The US cleaning process was done continuously until the microrods were dissolved.

Once the microvasculature was formed, MBs (SonoVue, Bracco, Imaging SpA, Milan, Italy) diluted as suggested by the manufacturer were injected in the phantom to verify the MB flow, as shown in figure \ref{fig:Methodology}f. The LA332 US probe and the ULA-OP 256 system were used to excite and acquire the US signals. The MB flow was visualised using the visual real-time feedback of the ULA-OP 256 system. The US imaging mode used was a coherent compounding of 3 angled plane waves (-$2.5^\circ$ to $2.5^\circ$) with a $3$ kHz pulse repetition frequency (PRF) and $19.5$ MHz sampling frequency. The excitation was a $3$ MHz, $3$-cycle, sine-weighted sinusoidal signal. 

\subsection{Reproducibility, Durability and Optical Validation}
\label{sec:Reproducibility_durability_optical}

The ground truth of the MC dimensions in the phantom was measured using the BX3M optical microscope (Evident, Olympus Corp., Japan). 

Two phantoms were fabricated to demonstrate the reproducibility of the methodology. One phantom contained a main vessel bifurcating into two secondary vessels, forming an angle of $2.8^\circ$, and a second phantom contained a main vessel branching into $2$ secondary vessels and $6$ tertiary vessels.  

The durability of the phantoms was evaluated one year after fabrication by assessing changes in weight, size, acoustic properties, and optical opacity, as well as checking for visible cracks or damage following fatigue flow and compression load tests. The fatigue flow load test involved $100$ cycles, each with water flowing through the phantoms for $6$ minutes at a rate of $150$ $\mu l/min$. The flow speed was selected as $2$ times the average blood flow velocity found in microvasculature \cite{ChenBloodCapillaryBloodFlow}. The flow speed was controlled with a syringe pump using a $1$ ml syringe. The flow load duration was selected as $2$ times the common acquisition time for SRUS \cite{ChristensenJeffriesSRUSimaging}. The number of cycles was based on the assumption that the phantom is used twice a week for a year (assuming $50$ weeks per year). The fatigue compression load test consisted of a $100$ N axial load repeated $500$ times on the phantoms. The axial load was the average pressure applied by humans handling objects \cite{BardoHumaGripFunctionalForce}. The number of cycles was selected assuming the phantom is handled ten times a week for a year (assuming $50$ weeks per year). The Instron machine $2519$ (Instron, Illinois Tool Works Inc., USA) was used to apply the axial load. The phantoms were considered durable if, after these tests, the sample's weight loss, size, acoustic properties and optical opacity changes were less than $10\%$, and there were no visible, under the microscope, cracks or damage in the phantoms; otherwise, the material was considered not durable and not suitable for the phantoms. The microvascular phantoms were stored at room temperature.

\section{Super-Resolution Ultrasound Imaging Methodology}
\label{sec:SRUS_imaging_methodology}

\subsection{Acquisition Strategy}

SonoVue MBs were diluted $1$ to $30$ with deionised water and administrated at a rate of $12$ $\mu$l/min using a syringe pump. A polytetrafluoroethylene (PTFE) microtube with $400$ $\mu$m internal diameter and $900$ $\mu$m external diameter (PTFE tubing, VWR International, LLC., UK) was connected between the syringe and the inlet of the phantom. The difference in diameters between the phantom inlet and the microtube provided good pressure to hold it in the phantom. A flat-tip needle was used to connect the microtube to the syringe. The flat tip prevented damage to the microtube when the needle was inserted into it. The syringe was frequently rotated to maintain an even distribution of the MB in the solution.

\begin{figure}[h]
\centering

        \includegraphics[width=0.9\linewidth]{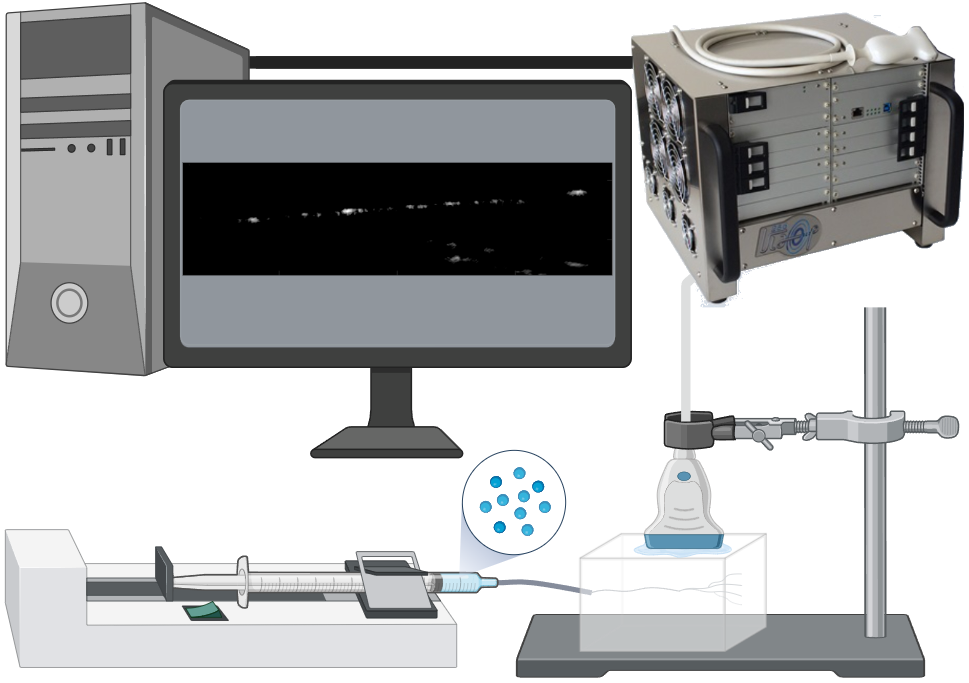}
   
\caption{Ultrasound imaging setup. The image was created with BioRender.com.}
\label{fig:US_setup}
\end{figure}

To acquire the US data, the LA332 US probe was placed on the surface of the phantom, as shown in figure \ref{fig:US_setup}. The probe was coupled to the phantom using ultrasonic gel. The probe is held in place with a holder, so motion is not induced during acquisition. The acquired US data was processed using Matlab in an Intel(R) Xeon(R) Gold 6226R CPU and 64 GB of RAM computer.

The radiofrequency (RF) signals were acquired using the ULA-OP 256 system. The total aperture of the transducer (144 elements) was utilised in transmission and reception. To image the phantoms, the same imaging modality described in section \ref{sec:phantom_fabrication_optical_validation} was utilised. US data to image $16$mm in depth per plane wave was collected. A total period of $5$ min of US data was recorded to image a phantom.

\subsection{Processing Pipeline}

The RF US signals were post-processed with sequential frame and temporal filters in batches of $1$ s. First, the mean of the frames was subtracted from the US data. This step eliminated non-moving US reflections from the MCs and the back wall of the phantom. Then, a low-pass frame filter and a band-pass temporal filter were implemented to improve the frames' signal-to-noise ratio.

A low-pass, $50$ Hz, frame filter was employed to eliminate any signal lasting less than $20$ ms per pixel. It was experimentally observed that noise between frames was faster than $20$ ms per pixel. In addition, when employing a lower cutoff, it was observed experimentally that MB signal amplitude was reduced.

A frequency filter with a band-pass of $1.3$ to $5.5$ MHz was implemented to eliminate offset and random noise in the RF signals. The band-pass filter preserved the frequency response of the excitation signal. No harmonic was observed in the frequency response of the RF signals.

The US data was then beamformed using delay-and-sum (DAS) and compressed in decibels (from -15 dB to 0 dB). The US data was beamformed with a pixel size of $50$ $\mu$m.

The SRUS processing started with localising potential MB signals in the filtered, beamformed and compressed US dataset. First, local maxima in each US frame were found using the Matlab function \textit{imregionalmax}. Then, the highest amplitude local maxima per frame were selected as potential MB signals. $10$ local maxima per frame were selected for the bifurcation phantom and $40$ for the microvascular phantom. These quantities were selected based on the number of MB signals per frame observed upon visual inspection of the US data.

The amplitude-centroiding method weighted average (WA) was used to estimate the precise location of the MB signals. WA is an amplitude-informed method that uses the amplitude distribution around a maximum to calculate its precise location. The area for the method was defined as $350$ $\mu$m by $350$ $\mu$m around the local maximum of the MB signal. The area selected was slightly smaller than the FWHM of the MB point spread function (PSF) to reduce processing time. For an MC with a diameter of $\sim60$ $\mu$m at a depth of $8$ mm, the experimentally observed FWHM of the MB PSF was $\sim400$ $\mu$m in the lateral direction. The weights of the WA method were selected as the pixel distance each pixel has in the lateral and depth direction to the maximum. Hence, the farther the pixel is, the greater the absolute value of the weight is. The location of the MB in pixels was estimated as shown in equation \ref{eq:weigthedaverage} \cite{HeilesSRUSPALABrainMouse}.

\begin{equation}
\begin{split}
l^* &=l_m+\frac{\sum_{j=-n}^{-n}\sum_{i=-n}^{n}{i*I(l_m+i,d_m+j)}}{\sum_{j=-n}^{-n}\sum_{i=-n}^{n}{I(l_m+i,d_m+j)}}, and \\
d^* &=d_m+\frac{\sum_{j=-n}^{-n}\sum_{i=-n}^{n}{i*I(l_m+j,d_m+i)}}{\sum_{j=-n}^{-n}\sum_{i=-n}^{n}{I(l_m+i,d_m+j)}}
\end{split}
\label{eq:weigthedaverage}
\end{equation}

where $(l^*,d^*)$ is the estimated location of the MB in pixels. $(l_m,d_m)$ is the position in pixels of the local maximum associated with the MB. $i$ and $j$ are the indexes of the pixels in the area. $I$ is the amplitude of the image. $n$ is the number of pixels in the lateral and depth direction of the area. For a $350$ $\mu$m by $350$ $\mu$m area with pixel size $50$ $\mu$m, $n$ is equal to $3$. The location of the MB in space can be calculated by multiplying its location in pixels with the pixel size.

A simple tracker based on the Euclidean distance was implemented to track MBs flowing in the microvasculature phantoms. MBs were linked if the Euclidean distance between two MBs was less or equal to $200$ $\mu$m and the frame separation was $1$ or $2$. The Euclidean distance was selected as half of the FWHM of the MB PSF to consider the variability of the observed MB signals. The frame separation was chosen as the minimum so only MB signals that continuously appear in the dataset were tracked. The linked MBs became a track if the number of linked MBs was or exceeded $15$ to ensure confidence in the tracked MB.        

The SRUS localisation maps were obtained by accumulating the tracked MBs into one dataset. The MC diameters were estimated by calculating the FWHM of the accumulated localisations at a given MC cross-section. The FWHM was calculated directly from the experimental results. When the half-magnitude values were unavailable, a Makima interpolation between the nearest neighbours was used to estimate the value.   

The microvasculature density map was calculated as the number of pixels where MBs were located divided by the total pixels in the query area. The global microvasculature density was calculated over the total imaging area of the microvasculature phantom ($35.35$x$8.55$ mm$^2$). The local microvasculature density was calculated over subset areas of $1$x$1$ mm$^2$.

For visualisation purposes only, the SRUS images were rendered by accumulating all the tracked MBs multiplied by a Gaussian distribution with a mean of $25$ and a standard deviation of $25$ in the lateral and depth directions before the summation. 

\subsection{Performance Evaluation}
\label{sec:SRUS_imaging_performance}

The trueness -how close a measurement is to its true value- and the precision -how close to each other are repeated measurements of the same quantity- \cite{ISO5725-1} were estimated to evaluate the performance of the SRUS results. The SRUS trueness and precision were calculated by comparing the SRUS estimated MC diameters with their optical measurements. The optical measurements were used as the true diameter values. Three MCs and $5000$ SRUS diameter measurements per MC were used to calculate its trueness and precision. The improvement of the SRUS trueness and the precision as the MBs localisation increased were estimated. 

\section{Results}
\label{sec:results}

\begin{figure}[h]
\centering

        \includegraphics[width=0.6\linewidth]{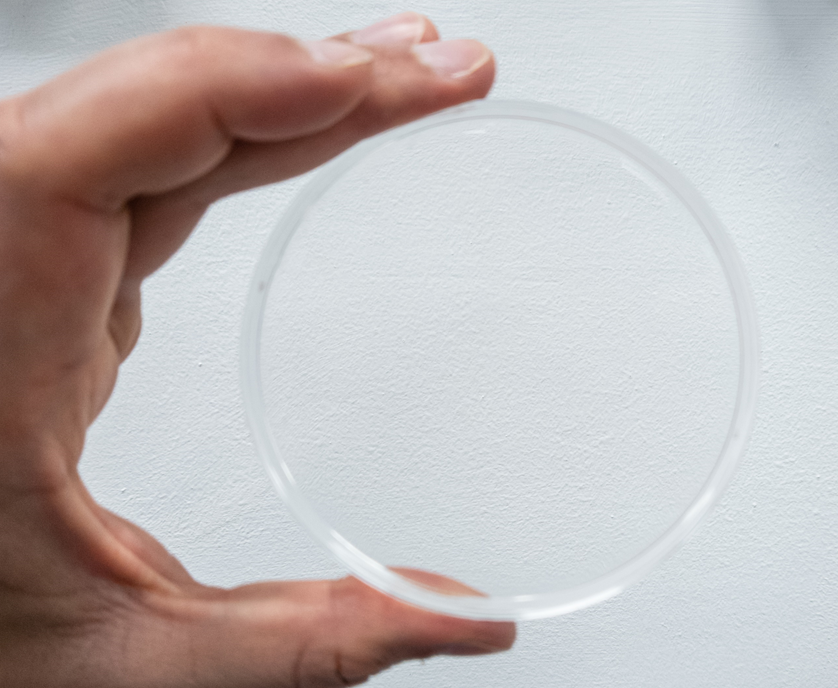}
        \label{fig:pdms}
        
\caption{Disc sample of PDMS tissue-mimicking material. }
\label{fig:TMM}
\end{figure}

Figure \ref{fig:TMM} shows a disc sample of the PDMS TMM, where its optical opacity can be observed. The photo was taken under the same light conditions as the optical opacity experiments.

Table \ref{tab:TMM} presents the acoustic and optical opacity properties of the TMM at $0$ year (after demolding) and $1$ year after fabrication. 

\begin{table}[h!]
\centering
\begin{tabular}[c|c|c|c]{p{1.8cm}|p{1.1cm}|p{2.1cm}|c}
 \multicolumn{4}{c}{\textbf{Properties of the Tissue Mimicking Material}} \\
 \toprule
 \multirow{3}{*}{\textbf{Material}} &  \centering \textbf{Wave} &  \centering \textbf{Acoustic} &  \textbf{Optical}  \\
                                    &  \centering \textbf{Velocity} &  \centering\textbf{Attenuation at} &  \textbf{Opacity}  \\
                                    & \centering \textbf{[m/s]} & \centering \textbf{$3$ MHz [dB/cm]} &\textbf{[\%]} \\
 \midrule
 \textbf{PDMS (0 year)} & \centering $1032.5$ & \centering $11.6$ & $8.7$ \\
 \midrule
 \textbf{PDMS (1 year)} & \centering $1044.2$ & \centering $10.6$ & $12.3$ \\
  \bottomrule
\end{tabular}
\caption{Acoustic properties and optical opacity of the tissue mimicking material.}
\label{tab:TMM}
\end{table}

Table \ref{tab:durable} presents the weight loss, size reduction, compression and flow fatigue tests, and the durability results of the phantoms $1$ year after demolding. 

\begin{table}[h!]
\centering
\begin{tabular}[c|c|c|c|c|c]{m{1.7cm}|m{0.9cm}|m{0.9cm}|m{0.9cm}|m{0.9cm}|c  }
 \multicolumn{6}{c}{\textbf{Durability of the Microvascular Phantom a Year After Fabrication}} \\
 \toprule
 {\multirow{3}{*}{\centering \textbf{Phantom}}} & \centering \textbf{Weight} & \centering \textbf{Size} & {\centering { \centering \textbf{Cyclic}}} & {\centering { \centering \textbf{Cyclic}}} & {\multirow{3}{*}{\centering \textbf{Durable}}}\\
             & \centering \textbf{Loss} & \centering \textbf{Loss} &  \centering{\textbf{Flow}} &  \centering{\textbf{Load}} &   \\
& \centering \textbf{[\%]} & \centering \textbf{[\%]} &  \centering{\textbf{Test}} &  \centering{\textbf{Test}} &   \\
 \midrule
 \textbf{Bifurcating} & \centering $2.4$ & \centering  $2.1$ & \centering {\centering Pass} & \centering {\centering Pass} & Yes\\
 \midrule
 \textbf{Microvascular} & \centering $3.1$ & \centering  $3.7$ & \centering {\centering Pass} & \centering {\centering Pass} & Yes\\
 \bottomrule
\end{tabular}
 \caption{Weight loss, size reduction, and durability test results of the bifurcating and microvascular phantom. The weight loss and size reduction percentage represents the weight and size changes that the phantoms had a year after fabrication. The cyclic flow and load tests were done on the phantoms a year after fabrication.}
\label{tab:durable}
\end{table}

The results shown in table \ref{tab:durable} and \ref{tab:TMM} were used to determine the durability of the sample. The durability of the microvascular phantom is shown in table \ref{tab:durable}.

\begin{figure}[h!]
\centering

    \begin{subfigure}[b!]{0.985\linewidth}
        \includegraphics[width=1\linewidth]{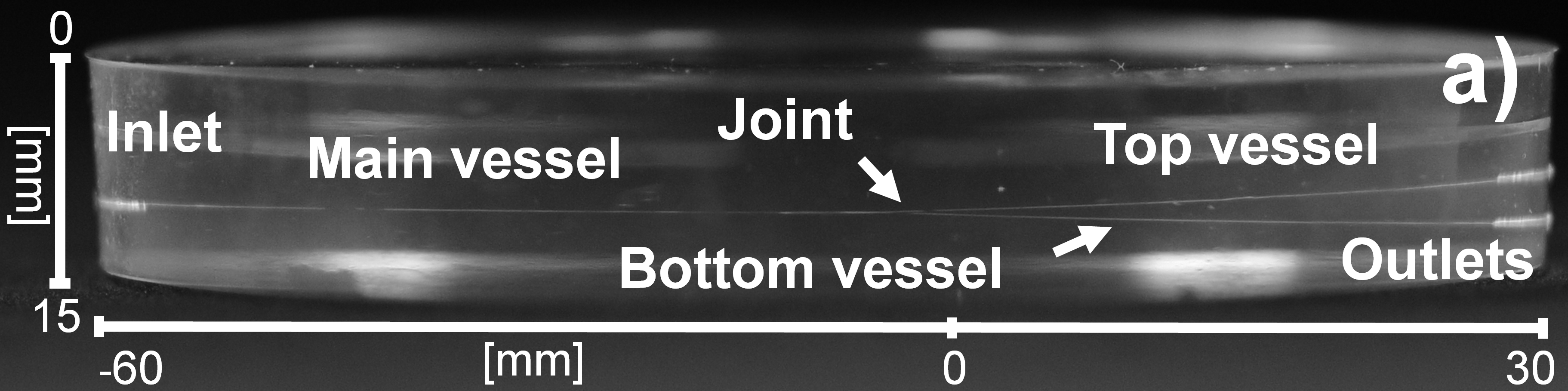}
        \label{fig:bifurcating_phantom_labels}
 \vspace{-0.3cm}
    \end{subfigure}

    \begin{subfigure}[b!]{0.48\linewidth}
        \includegraphics[width=1\linewidth]{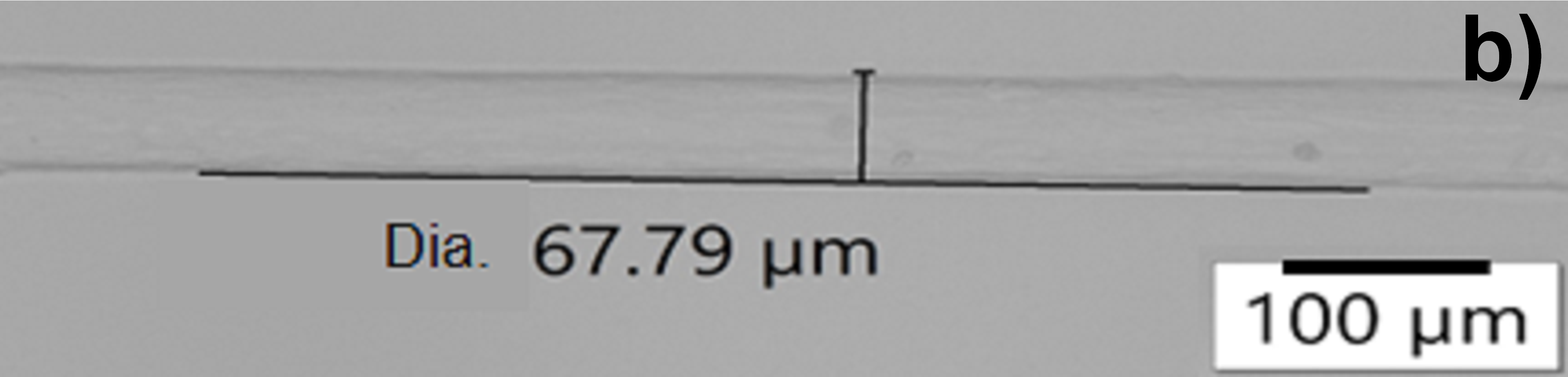}
        \label{fig:bifurcating_phantom_main_vessel}
 \vspace{-0.3cm}
    \end{subfigure}%
   ~
    \begin{subfigure}[b!]{0.48\linewidth}
        \centering
        \includegraphics[width=1\linewidth]{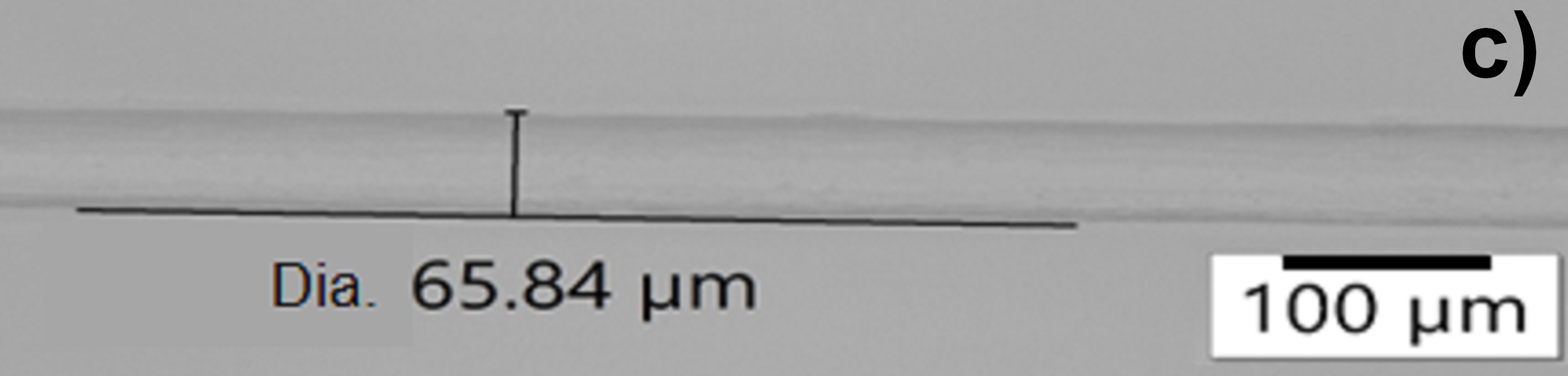}
        \label{fig:bifurcating_phantom_top_vessel}
 \vspace{-0.3cm}
    \end{subfigure}%

    \begin{subfigure}[b!]{0.48\linewidth}
        \centering
        \includegraphics[width=1\linewidth]{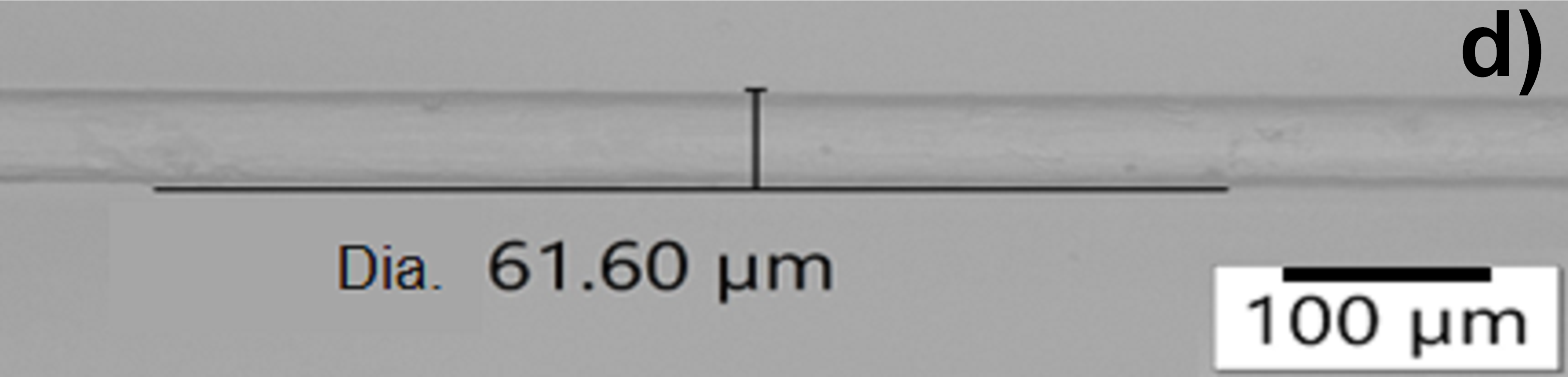}
        \label{fig:bifurcating_phantom_bottom_vessel}
    \end{subfigure}%
      ~
    \begin{subfigure}[b!]{0.48\linewidth}
        \centering
        \includegraphics[width=1\linewidth]{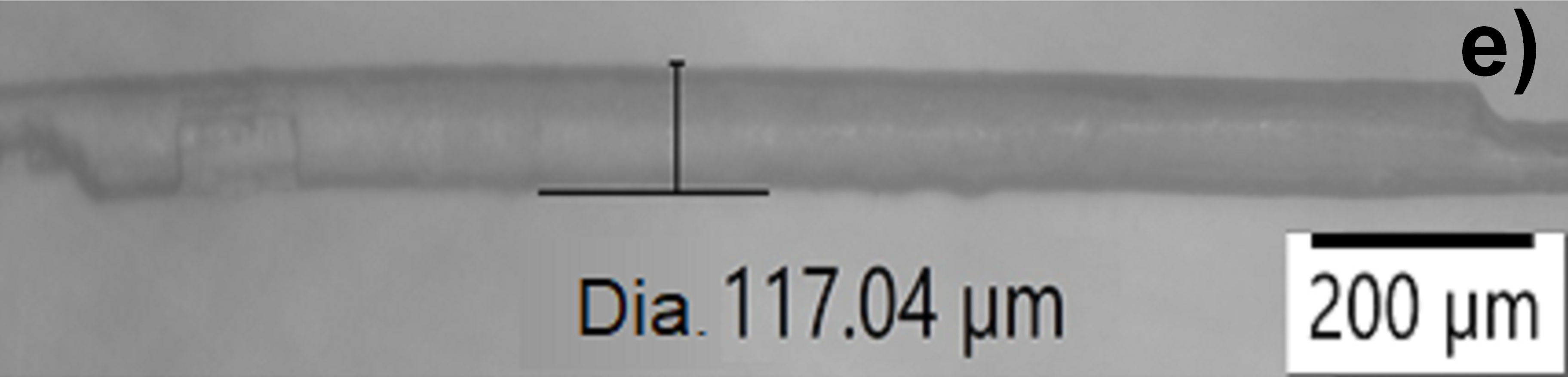}
        \label{fig:bifurcating_phantom_joint}
    \end{subfigure}%
        
\caption{Bifurcating microchannel phantom in a PDMS supporting matrix. a) Bifurcating phantom. b) Optical image of the main vessel. c) Optical image of the top vessel. d) Optical image of the bottom vessel. e) Optical image of the bifurcation joint. Optical images shown in b), c), d) and e) were taken with a microscope. The microscope used a top LED (white-balanced) light due to the shallow depth of the vessels. }
\label{fig:bifurcating_phantom}
\end{figure}

A bifurcating MC phantom was manufactured following the methodology presented in section \ref{sec:phantom_fabrication_optical_validation}. The bifurcating phantom comprised an MC -the main vessel- that bifurcated into two separate MCs -the top and bottom vessels- as shown in figure \ref{fig:bifurcating_phantom}a. The phantom was moulded using a petri dish and extracted after the PDMS was cured. The main vessel was connected to an inlet, and the top and bottom vessels were connected to outlets to let the MB flow in the MCs. The vessels' diameters were measured using the BX3M optical microscope. The optically measured diameters were $67.79$ $\mu$m, $65.84$ $\mu$m, and $61.60$ $\mu$m for the main, top, and bottom vessels, respectively. The optically measured length of the bifurcating joint was $1.35$ mm in the lateral direction. The optical measurements are shown in figure \ref{fig:bifurcating_phantom}. Measurement of MC sizes using the microscope demonstrated that diameters remained unchanged, both with and without flow.

\begin{figure}[h!]
\centering

    \begin{subfigure}[b!]{0.985\linewidth}
        \includegraphics[width=1\linewidth]{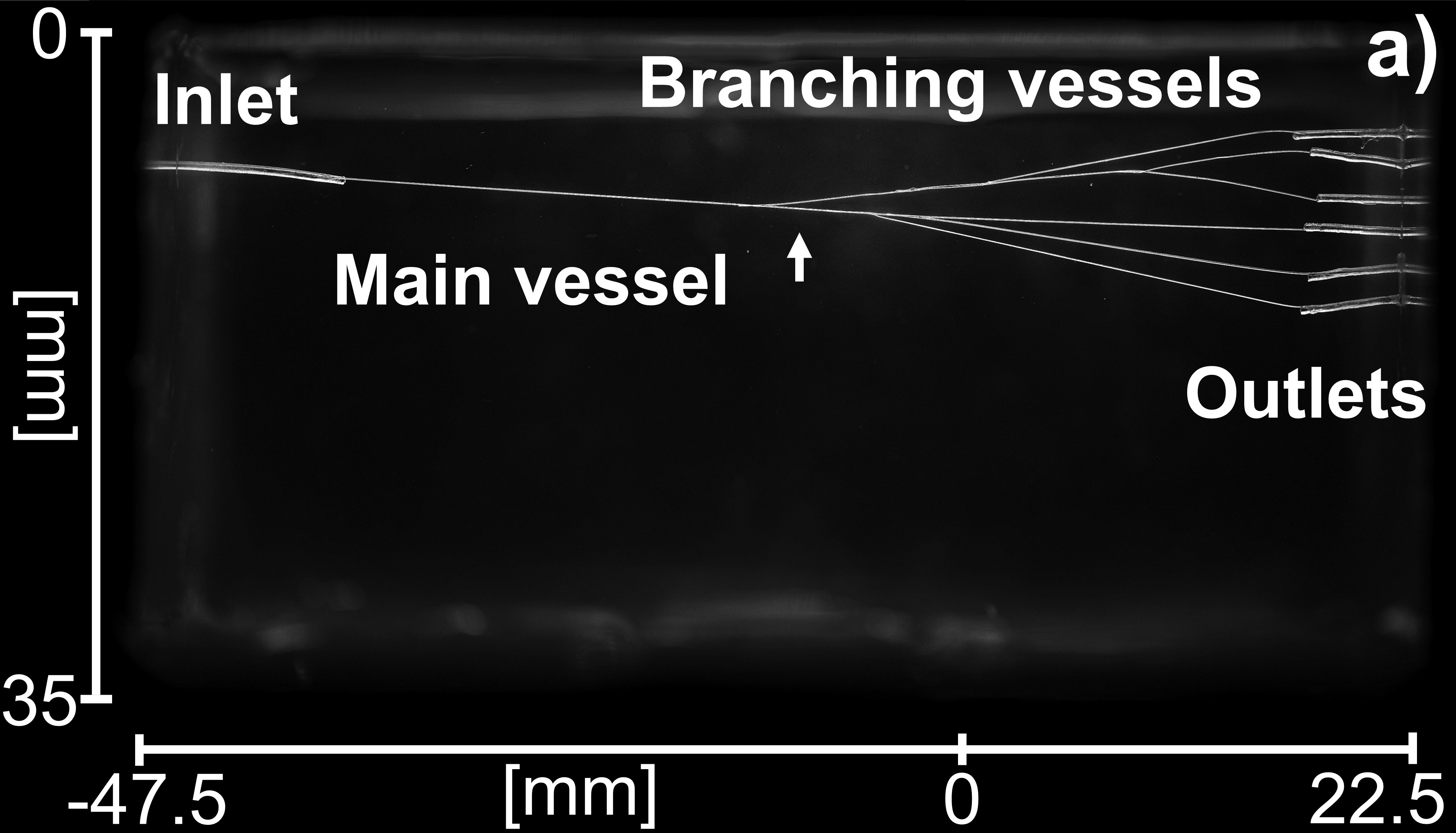}
        \label{fig:ramification_phantom_labels}
 \vspace{-0.3cm}
    \end{subfigure}

    \begin{subfigure}[b!]{0.48\linewidth}
        \includegraphics[width=1\linewidth]{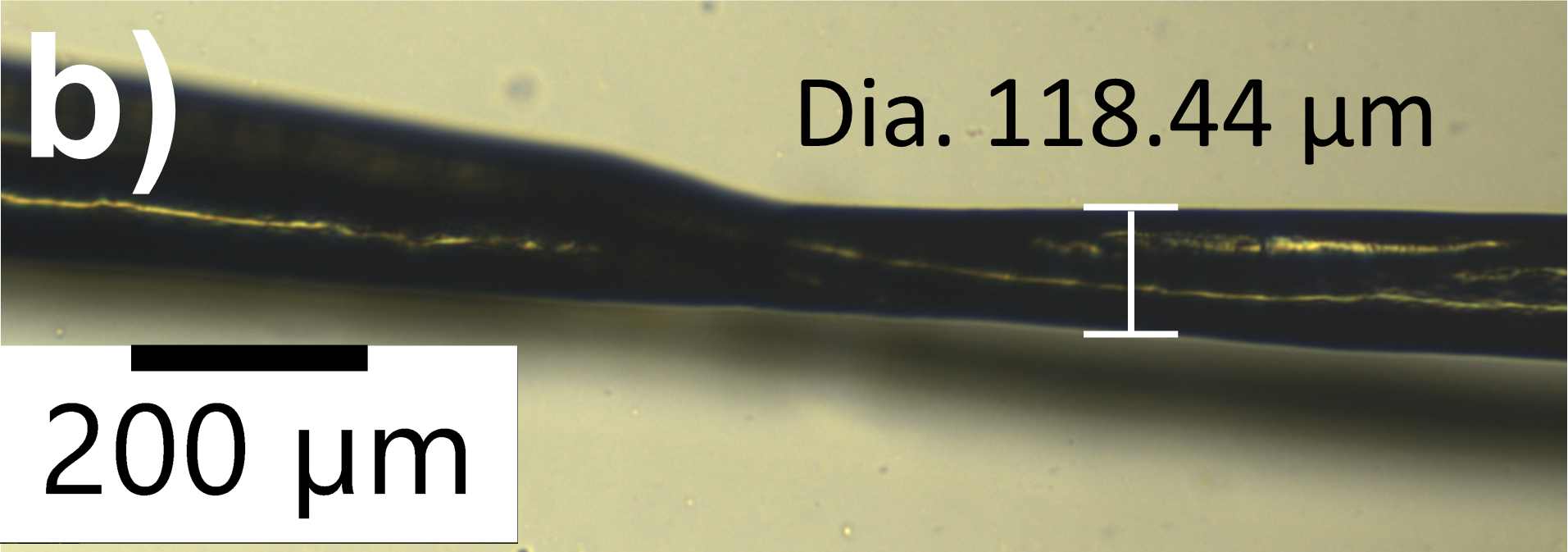}
        \label{fig:ramification_phantom_top_secondary_vessel}
 \vspace{-0.3cm}
    \end{subfigure}%
   ~
    \begin{subfigure}[b!]{0.48\linewidth}
        \centering
        \includegraphics[width=1\linewidth]{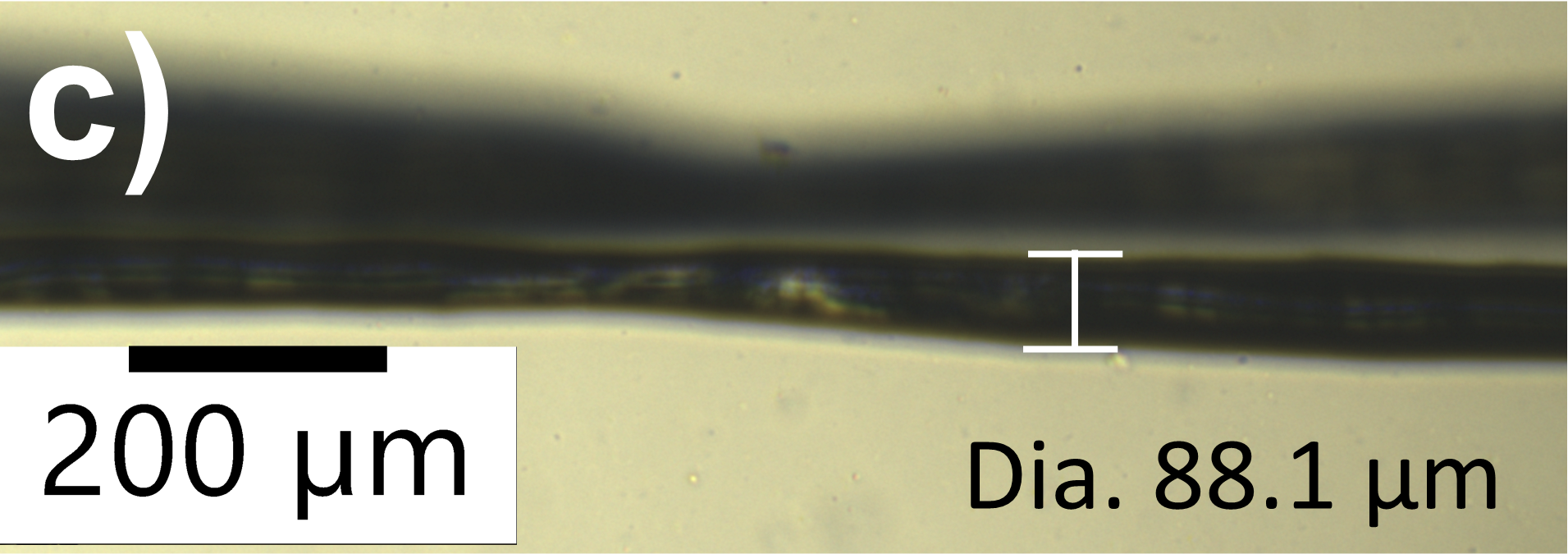}
        \label{fig:ramification_phantom_bottom_secondary_vessel}
 \vspace{-0.3cm}
    \end{subfigure}%

\caption{Microvascular phantom in a PDMS supporting matrix. a) Microvascular phantom. b) Optical image of the top secondary vessel. The white lines in the vessel are reflections of the incident light. The black, out-of-focus area corresponds to a vessel in the distance. c) Optical image of the bottom secondary vessel.  The black, out-of-focus area corresponds to a vessel in the distance. The white arrow in image a) indicates where the microscope images shown in b) and c) were taken. The microscope used a bottom incandescent (yellow) light due to the depth of the vessels.}
\label{fig:microvasculature_phantom}
\end{figure}

Similarly to the bifurcating phantom, a microvascular phantom was manufactured using the methodology described in section \ref{sec:phantom_fabrication_optical_validation}. The microvascular phantom was composed of a main vessel branching into several vessels, as shown in figure \ref{fig:microvasculature_phantom}a. The main vessel bifurcated into two secondary vessels, and each secondary vessel branched into three tertiary vessels following the main microvascular structure of the ear phantom presented in \cite{ChristensenJeffriesSRUSimagingEarMouse}. Figure \ref{fig:microvasculature_phantom}b and \ref{fig:microvasculature_phantom}c present optical images taken with a microscope of the top and bottom secondary vessels, respectively.

\begin{figure}[h!]
\centering

    \begin{subfigure}[b!]{1\linewidth}
      \centering
         \includegraphics[width=1\linewidth]{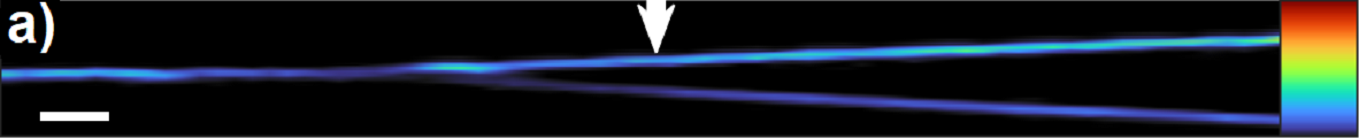}
        \label{fig:SRUS_image_bifurcating_phantom}
 \vspace{-0.25cm}
    \end{subfigure}

    \begin{subfigure}[b!]{1\linewidth}
        \centering
        \includegraphics[width=1\linewidth]{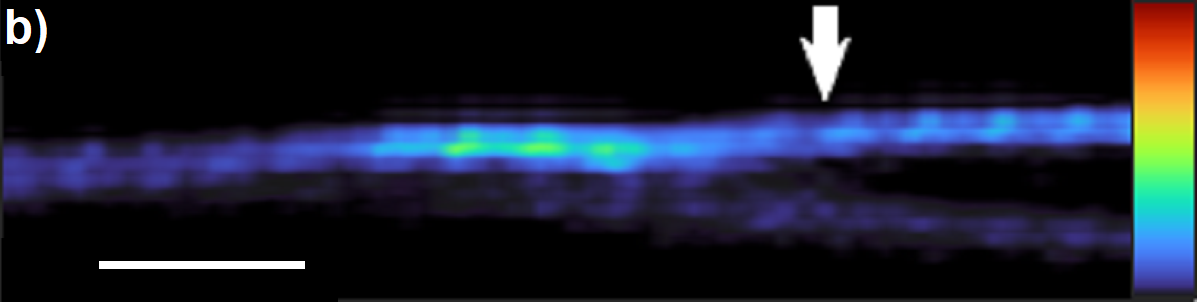}
        \label{fig:SRUS_zoom_joint}
 \vspace{-0.25cm}
    \end{subfigure}%

    \begin{subfigure}[b!]{0.48\linewidth}
        \centering
        \includegraphics[width=1\linewidth]{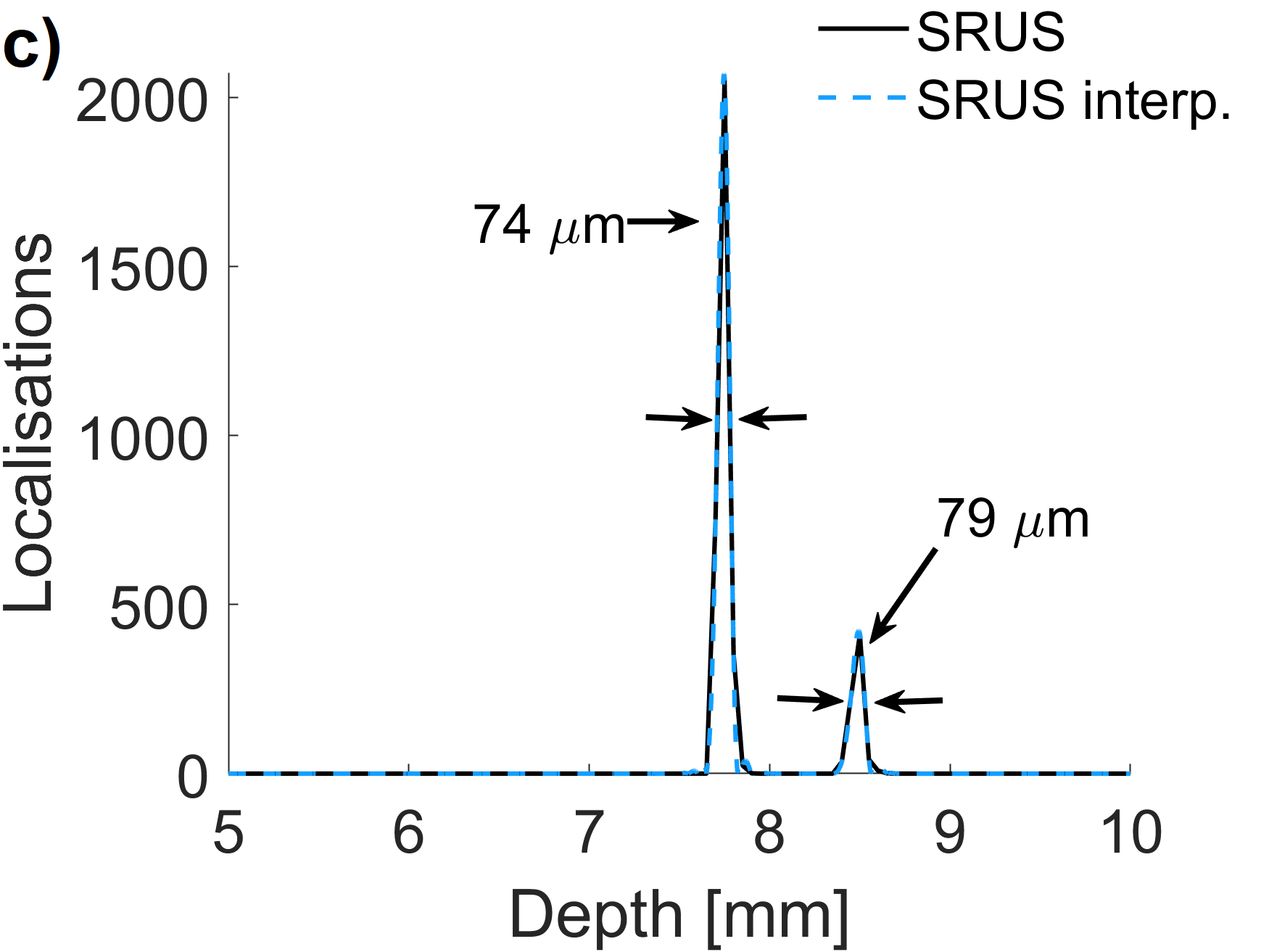}
        \label{fig:SRUS_zoom_joint}
 \vspace{-0.25cm}
    \end{subfigure}%
~
    \begin{subfigure}[b!]{0.48\linewidth}
      \centering
         \includegraphics[width=1\linewidth]{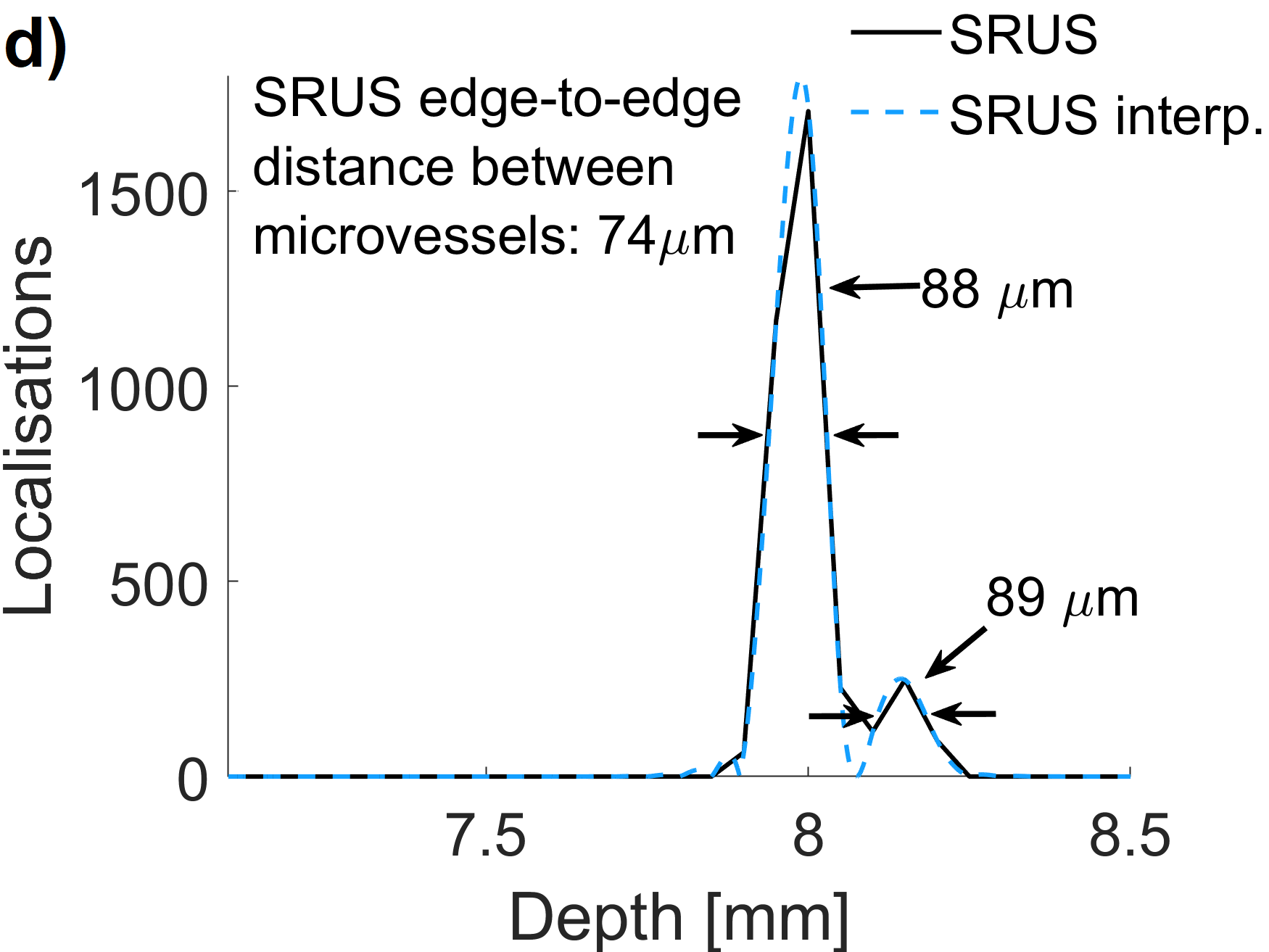}
        \label{fig:SRUS_image_bifurcating_phantom_line}
 \vspace{-0.25cm}
    \end{subfigure}
         
\caption{Super-resolution ultrasound results on the bifurcating phantom. a) SRUS image of the bifurcating phantom. Colour bar from 0 to 3500 localisations. b)  SRUS image at the joint area of the bifurcating phantom. Colour bar from 0 to 2500 localisations. Imaging depth 7 to 9 mm. Scale bar: 1mm. c) SRUS line plot illustrating the localisations on each bifurcating vessel. The location of the line is shown with a white arrow in a). d) SRUS line plot at the last location where the bifurcating vessels are distinguishable. The location of the line is shown with a white arrow in b). The black and cyan lines in c) and d) correspond to the SRUS results and interpolated SRUS results, respectively. The horizontal black arrows in c) and d) indicate the location of the full-width-half-magnitude of each vessel. }
\label{fig:SRUS_bifurcating_phantom}
\end{figure} 

SRUS results of the bifurcating phantom are shown in figure \ref{fig:SRUS_bifurcating_phantom}. Figure \ref{fig:SRUS_bifurcating_phantom}a shows the SRUS image of the bifurcating phantom, and figure \ref{fig:SRUS_bifurcating_phantom}b presents a zoom on the bifurcation joint. Figure \ref{fig:SRUS_bifurcating_phantom}c and \ref{fig:SRUS_bifurcating_phantom}d present a line plot of the SRUS results across the depth of the imaging region. The white arrow in figure \ref{fig:SRUS_bifurcating_phantom}a and \ref{fig:SRUS_bifurcating_phantom}b indicates the corresponding location of the line plots in figure \ref{fig:SRUS_bifurcating_phantom}c and \ref{fig:SRUS_bifurcating_phantom}d, respectively. The SRUS image in figure \ref{fig:SRUS_bifurcating_phantom}b shows that the joint region of the three channels has an extension of approximately $1.572$ mm in the horizontal direction. This distance was measured from the start of the joint until it bifurcated. The horizontal arrows in figures \ref{fig:SRUS_bifurcating_phantom}c and \ref{fig:SRUS_bifurcating_phantom}d indicate the location of the FWHM. Figure \ref{fig:SRUS_bifurcating_phantom}d corresponds to the last point in the lateral direction where the joint was separable in the SRUS results. The SRUS measured separation between the bifurcating vessels was $74$ $\mu$m. The distance was measured between the closest edges of each vessel.

\begin{figure}[h!]
\centering

 \begin{subfigure}[b!]{1\linewidth}
        \centering
        \includegraphics[width=1\linewidth]{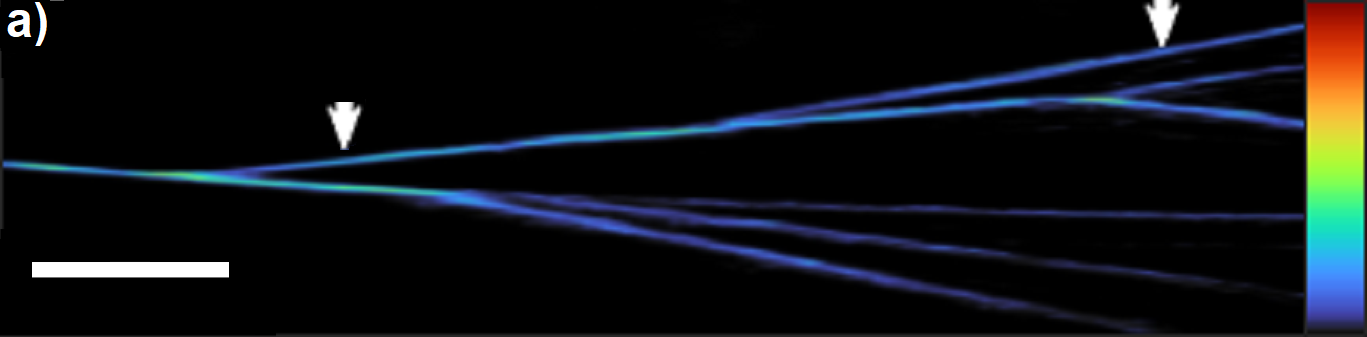}
    \vspace{-0.25cm}
    \end{subfigure}%

 \begin{subfigure}[b!]{0.48\linewidth}
      \centering
         \includegraphics[width=1\linewidth]{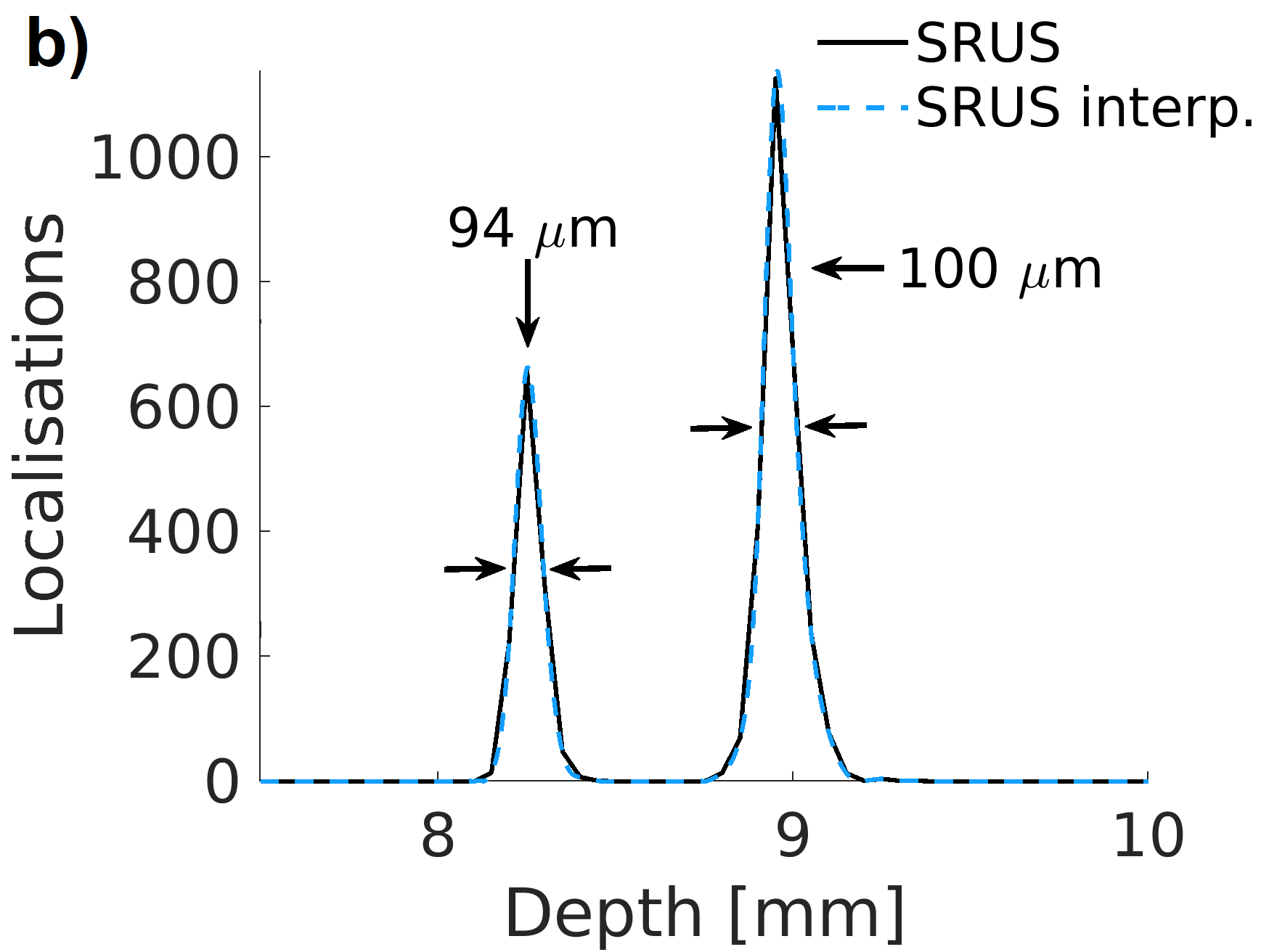}
        \label{fig:SRUS_image_ramification_phantom_line2}
 \vspace{-0.25cm}
    \end{subfigure}
~
    \begin{subfigure}[b!]{0.48\linewidth}
      \centering
         \includegraphics[width=1\linewidth]{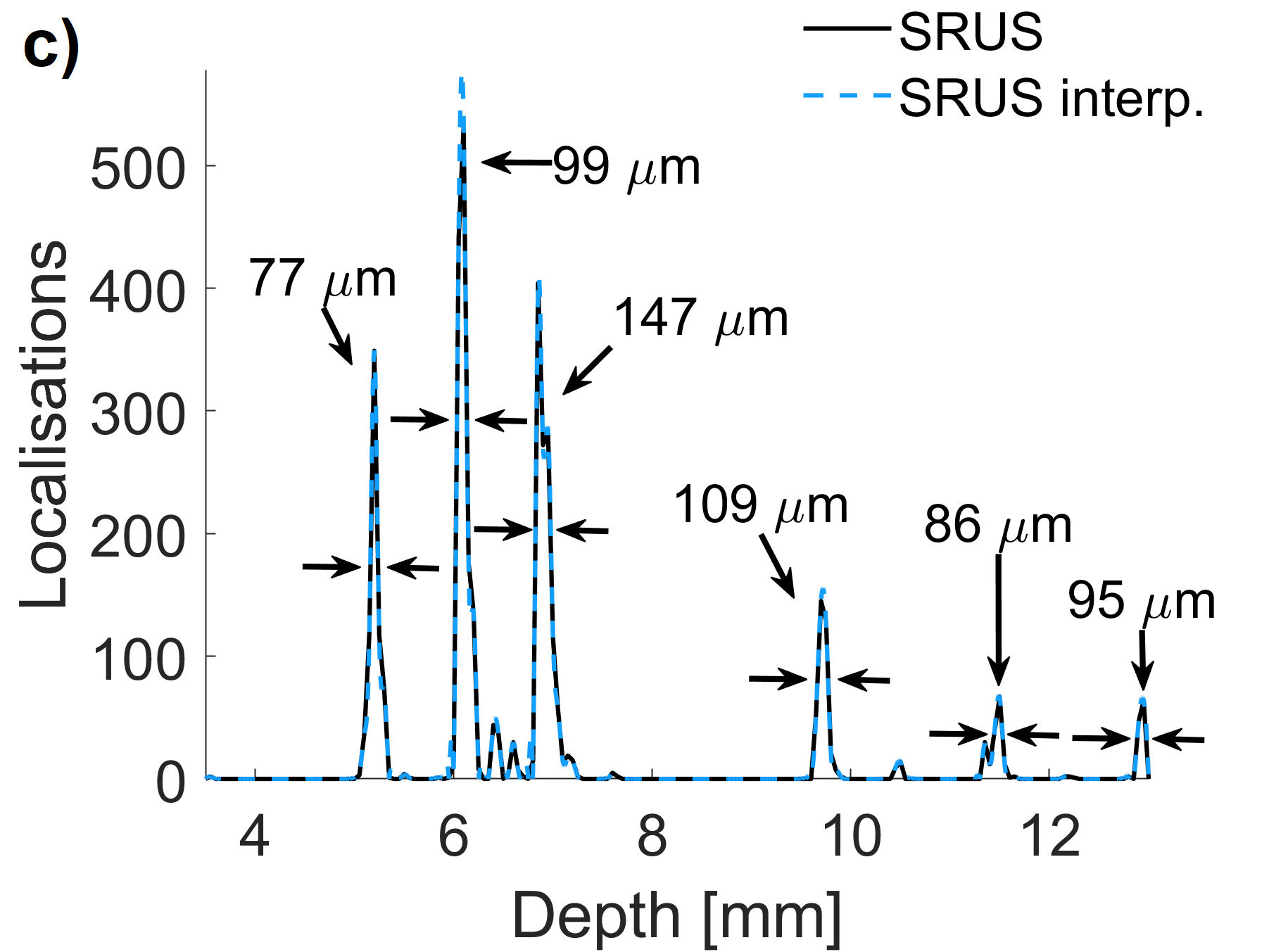}
        \label{fig:SRUS_image_ramification_phantom_line}
 \vspace{-0.25cm}
    \end{subfigure}

\caption{Super-resolution ultrasound results on the microvascular phantom. a) SRUS image of the microvascular phantom. Colour bar from 0 to 2500 localisation. Imaging depth 4 to 12.9 mm. Scale bar: 5mm. b) SRUS line plot illustrating the localisations obtained on the first bifurcation vessels. The location of the line is shown with the left white arrow in a). c) SRUS line plot illustrating the localisations obtained on the tertiary vessels. The location of the line is shown with the right white arrow in a). The black and cyan lines correspond to the SRUS and interpolated SRUS results, respectively. The horizontal black arrows in b) and c) indicate the location of the full-width-half-magnitude of each vessel.}
\label{fig:SRUS_microvasculature_phantom}
\end{figure}

SRUS results of the microvascular phantom are shown in figure \ref{fig:SRUS_microvasculature_phantom}. Figure \ref{fig:SRUS_microvasculature_phantom}b and \ref{fig:SRUS_microvasculature_phantom}c show a line plot of the obtained SRUS results on the microvascular phantom. The line plots correspond to the localisations along the depths indicated with white arrows in figure \ref{fig:SRUS_microvasculature_phantom}a. Figure \ref{fig:SRUS_microvasculature_phantom}b and  \ref{fig:SRUS_microvasculature_phantom}c correspond to the white arrows on the left and right, respectively. The location of the lines was selected to present the localisations across the most vessels possible. The error of the SRUS estimated diameters in figure \ref{fig:SRUS_microvasculature_phantom}b was calculated by comparing them with their microscope images shown in figure \ref{fig:microvasculature_phantom}b and \ref{fig:microvasculature_phantom}c. The calculated SRUS error was $18.44$ $\mu$m and $5.9$ $\mu$m for the top and bottom secondary vessel, respectively.

\begin{figure}[h!]
\centering

    \begin{subfigure}[b!]{0.95\linewidth}
      \centering
         \includegraphics[width=1\linewidth]{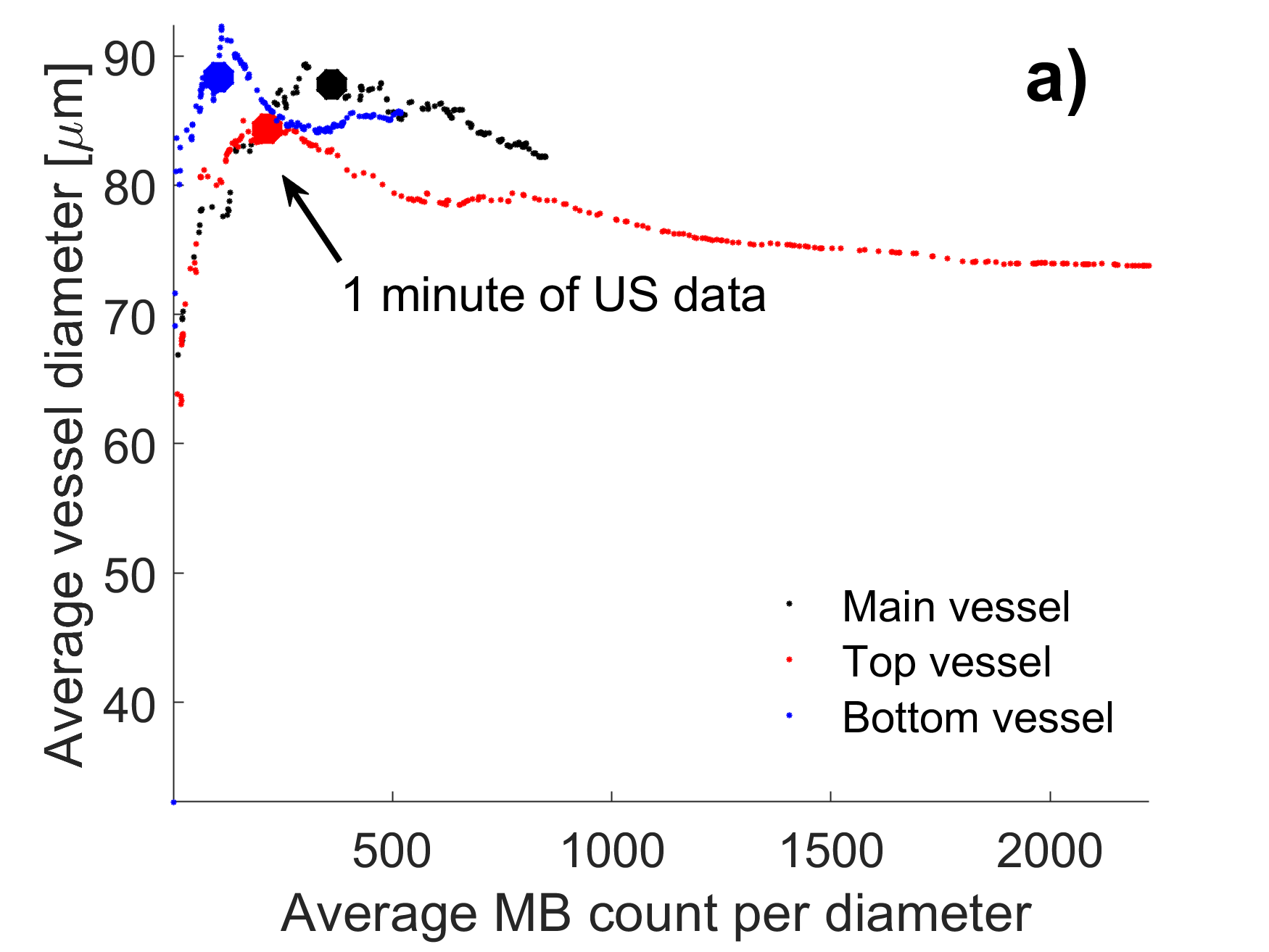}
    \end{subfigure}

    \begin{subfigure}[b!]{0.95\linewidth}
        \centering
        \includegraphics[width=1\linewidth]{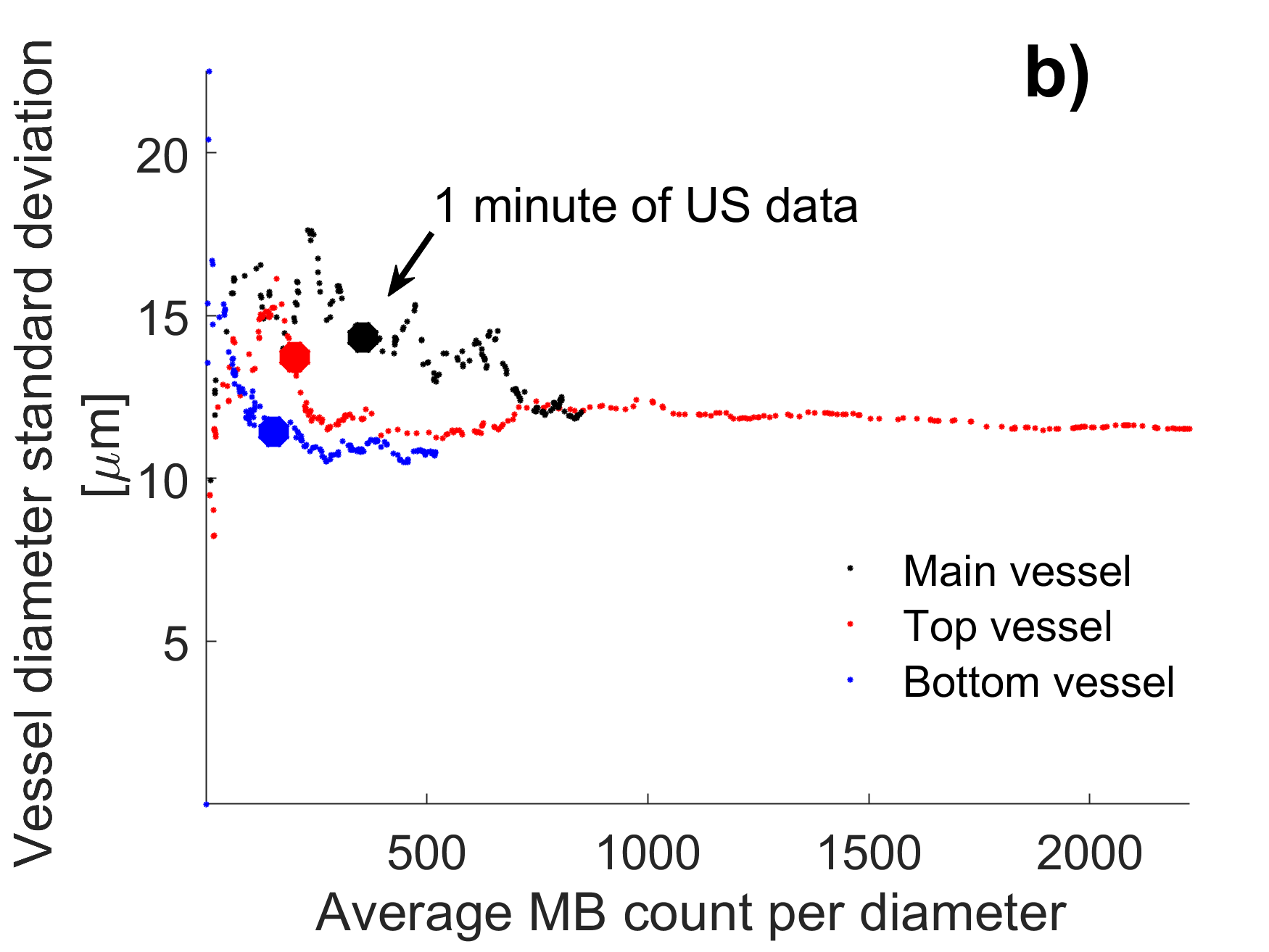}
    \end{subfigure}%
         
\caption{Estimated vessel diameters of the bifurcating phantom as a function of the average microbubble count per estimated diameter. The vessel diameters were calculated from the SRUS results. The average diameter and its standard deviation were calculated on a $5$ mm section per vessel. a) Average diameter per vessel. b) Standard deviation of the estimated diameters.}
\label{fig:SRUS_performance_stats}
\end{figure}

The diameters of the bifurcating phantom vessels were estimated from the SRUS results and compared against the optical measurements. A $5$ mm section per vessel was used to estimate the diameters. Hence, a total of $15000$ diameters were calculated from the SRUS results. On average, the estimated diameters were $82.20$ $\mu$m, $73.77$ $\mu$m, and $85.65$ $\mu$m for the main, top, and bottom vessels, respectively. The calculated average errors were $14.86$ $\mu$m, $7.93$ $\mu$m, and $24.05$ $\mu$m for the bifurcating phantom main vessel, top vessel, and bottom vessel. Similarly, the standard deviation of the estimated diameters were $12.02$ $\mu$m, $11.52$ $\mu$m, and $10.79$ $\mu$m for the main, top, and bottom vessels, respectively. The trueness and the precision of the SRUS methodology implemented were $15.44$ $\mu$m and $11.44$ $\mu$m. Figure \ref{fig:SRUS_performance_stats} presents the average estimated diameter and its standard deviation as a function of the average MB localised per estimated diameter. The values in figure \ref{fig:SRUS_performance_stats} were updated every second of US data processed. Hence, two consecutive points in figure \ref{fig:SRUS_performance_stats} were calculated from two sequential seconds in the US dataset.

\begin{figure}[h!]
\centering

    \begin{subfigure}[b!]{1\linewidth}
      \centering
         \includegraphics[width=1\linewidth]{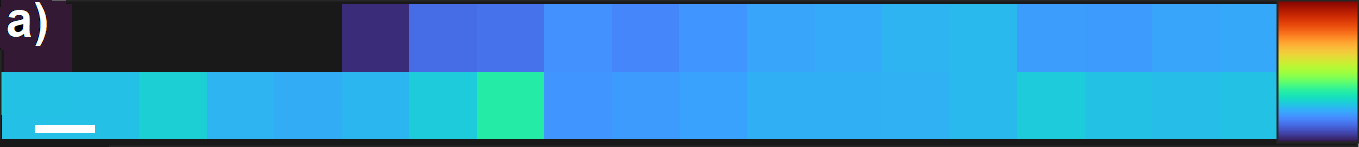}
        \label{fig:SRUS_LMVD_bifurcating_phantom}
 \vspace{-0.25cm}
    \end{subfigure}
  
    \begin{subfigure}[b!]{1\linewidth}
        \centering
        \includegraphics[width=1\linewidth]{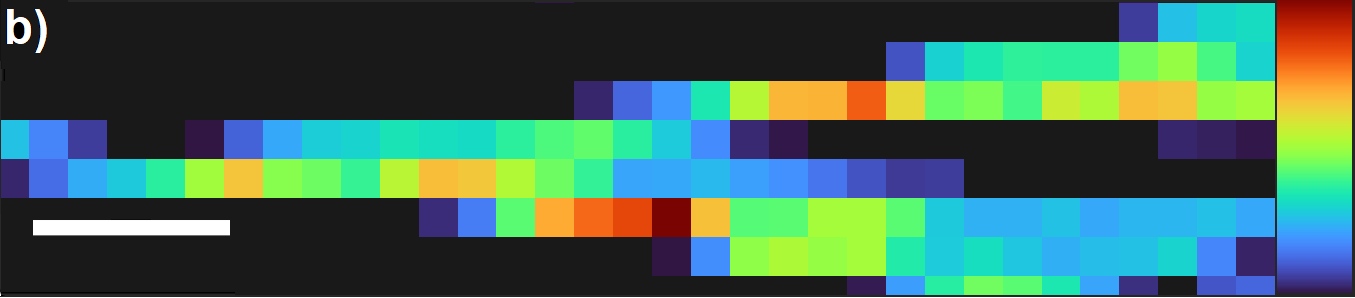}
        \label{fig:SRUS_LMVS_ramification_phantom}
\vspace{-0.25cm}
    \end{subfigure}%

  \begin{subfigure}[b!]{0.48\linewidth}
        \includegraphics[width=1\linewidth]{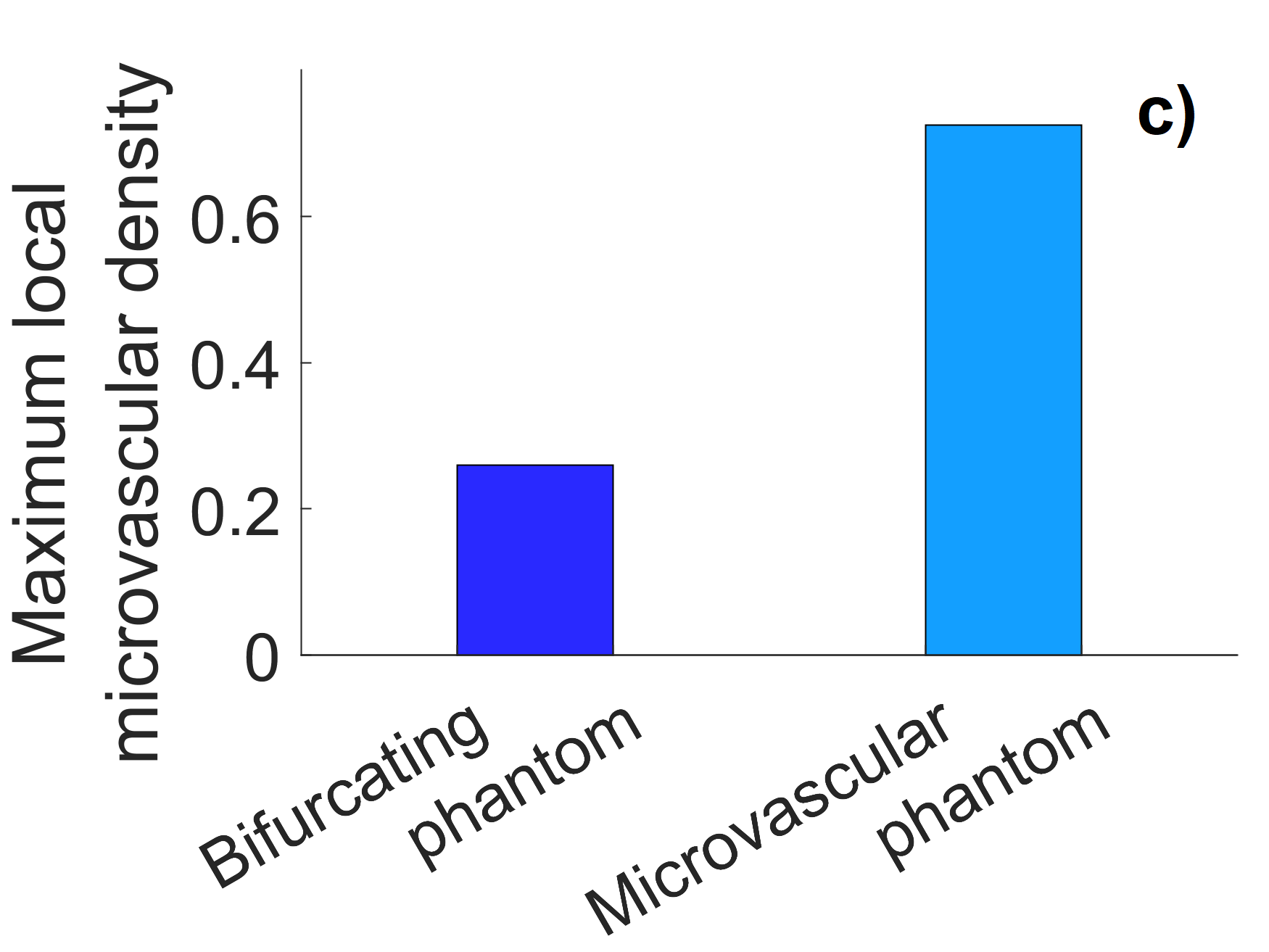}
        \label{fig:SRUS_GMVD_comparison}
 \vspace{-0.3cm}
    \end{subfigure}%
   ~
    \begin{subfigure}[b!]{0.48\linewidth}
        \centering
        \includegraphics[width=1\linewidth]{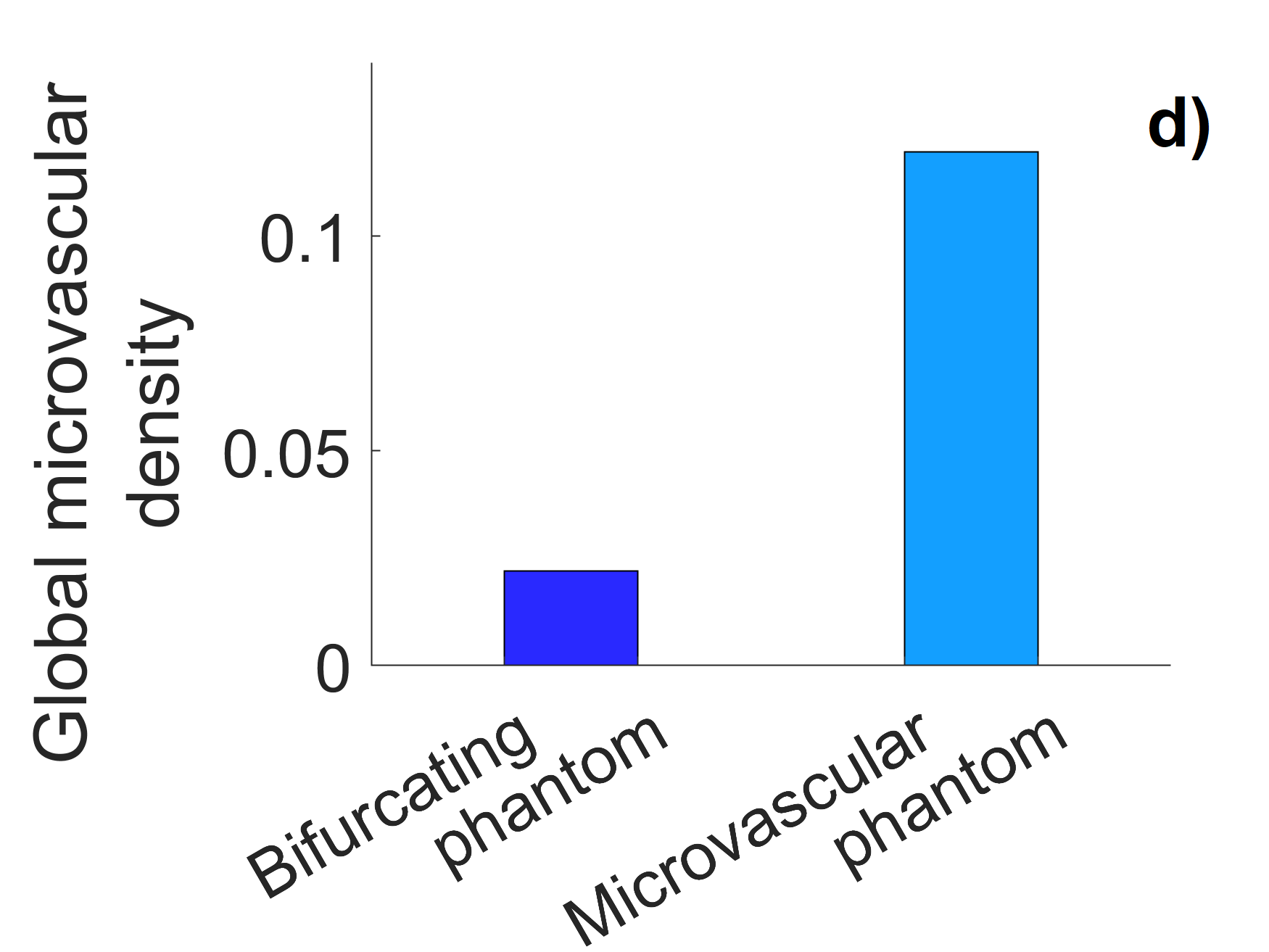}
        \label{fig:SRUS_GMVD_comparison}
 \vspace{-0.3cm}
    \end{subfigure}%
         
\caption{Microvascular density of the bifurcating and microvascular phantoms. a) Local microvascular density of the bifurcating phantom. White scale bar: 1mm. b) Local microvascular density of the ramification phantom. White scale bar: 5mm. The colour bar in a) and b) goes from $0$ to $72.5\%$. The local microvascular density was calculated on an area of $1$x$1$ mm$^2$. c) Maximum local microvascular density of the phantoms. d) Global microvascular density of the phantoms. The global microvascular density was calculated on the imaging area ($35.35$x$8.55$ mm$^2$).}
\label{fig:SRUS_GLMVD}
\end{figure}

The local microvascular density of the bifurcating and microvascular phantoms are shown in figures \ref{fig:SRUS_GLMVD}a and \ref{fig:SRUS_GLMVD}b. The local microvascular density was calculated on a local area of $1$x$1$ mm$^2$. Figure \ref{fig:SRUS_GLMVD}c presents the maximum local microvascular density for each phantom. The global microvascular density of each phantom is presented in figure \ref{fig:SRUS_GLMVD}d. The vascular variance of the bifurcating and microvascular phantom was $0.026$ and $0.105$, respectively. The global microvasculature density and its variance were calculated on the same area for both phantoms. The area used for the global microvascular density estimation was $35.35$x$8.55$ mm$^2$.

\section{Discussion}
\label{sec:discussion}

The PDMS TMM provided a suitable supporting matrix for the microvascular phantom. PDMS is easy to fabricate, stable at room temperature, durable, withstands the phantom manufacturing process, and is optically transparent. However, its wave speed is $\sim2/3$ of typical in vivo values, and its amplitude attenuation is $\sim5$ higher than soft tissue. Hence, the amplitude attenuation observed in the imaged microvasculature is equivalent to $\sim5$ cm deep in soft tissue.

Exploring the use of other TMMs and their suitability as a supporting matrix for the microvascular phantom could lead to phantoms that mimic specific parts of the body. For example, a TMM whose acoustic properties can be modified could be used to fabricate aberrating phantoms that mimic the different layers of human tissue or a specific organ. Similarly, finding a TMM that is optically transparent, mimics the acoustic properties of in vivo and is suitable for the proposed methodology can be used to study MB dynamics, where US-activated MBs are imaged with a high-speed camera.

Similarly, the fabrication methodology presented here can be combined with other methods to obtain phantoms that model the micro and macro dynamics of blood flow in the human body. For example, microvascular structures can be added to a heart phantom to investigate the performance of US imaging techniques to visualise the MCs while there is a high dynamic flow.

Ex vivo parts in the microvascular phantom can enhance the phantom tissue-mimicking properties. For example, a skull could be used to evaluate and improve US imaging algorithms designed to image the microvasculature of the brain. 

The laser-cut mould allowed easy alignment of the microrods. The mould contained laser-cut circular holes of $0.7$ mm in diameter where the inserts were placed. The precise cut of the mould would facilitate the repeatability of the phantom.

Microrods with greater diameters are easier to dissolve, especially when both ends are exposed to the solvent. Hence, when using the proposed methodology for the first time, it is recommended to use microrods a few hundred micrometres in diameter so fast setup adjustments can be implemented. Once the setup is ready, the method presented here is reliable, repeatable, and expandable to other US applications, which the authors are currently exploring.

The fabrication of microrods could be further improved. Currently, the microrods are fabricated by pulling ABS while extruding in a 3D printer. With this method, microrods smaller than $\sim50$ $\mu$m are difficult to achieve. However, this process could be automated by programming the 3D printer to pull the microrod while extruding. With this, a systematic method to fabricate the microrods with smaller diameters could be developed.

A different microrod material could be used to form the microvasculature negative print. This would allow the exploration of other solvents and more desirable TMMs in the proposed fabrication methodology. A solvent that does not evaporate at room temperature would facilitate the dissolution of the negative print, and a TMM with better US properties would allow the fabrication of a microvascular phantom with deeper microvascular features. The authors are currently working on this.

The design of the microvascular phantom included the general branching behaviour of a real in vivo microvasculature. In the future, other in vivo characteristics can be included, e.g., vessel size, $3$D spatial distribution, tortuosity and flow velocity, which would allow performance evaluation of other specific algorithms of the implemented SRUS strategy. 

It was observed that, sometimes, the MB flow did not flow in all the vessels at the same time during the experiments. This could be caused by the similar vessel diameters used across the phantoms. Hence, the MB flow in the phantoms could be improved by reducing the vessel diameters as a vessel branch. This could be added to the phantom design using the fabrication methodology presented here. The authors will explore this in future work.  

While images shown in figures \ref{fig:bifurcating_phantom}b-e and in figures \ref{fig:microvasculature_phantom}b-c were taken with the same microscope, the contrast between the vessels and the TMM matrix is different between figures. The reason for this is the light used to illuminate the vessels. In figures \ref{fig:bifurcating_phantom}b-e, a white-balanced LED light from the top of the microscope was used. While in figures \ref{fig:microvasculature_phantom}b-c, yellow incandescent light from the bottom of the microscope was used. Top light was selected to visualise shallow vessels due to its superior vessel edge contrast, clarity and no colour cast, while bottom light was selected to visualise vessels that were not visible with top light due to their depth in the phantom.

It was observed that the maximum average error of $24.05$ $\mu$m ($\sim\lambda/14$) occurred when the average MB count per diameter in the bottom vessel of the bifurcating was $519.3$. In contrast, the minimum average error ($7.93$ $\mu$m, $\sim\lambda/44$) occurred when the average MB count per diameter in the top vessel was $2224.82$, suggesting that the maximum error could be reduced if more MBs were localised in the bottom vessel. The average error of $7.93$ $\mu$m ($\sim\lambda/44$) and its standard deviation of $11.52$ $\mu$m show improved results compared to the predicted theoretical localisation error in \cite{DesaillySRUSLocalisationErrorPDMS}.

The average error and the standard deviation of the SRUS results were computed as the average MB count per diameter increased. It was observed that the SRUS algorithm would overestimate the diameters of the microvessels when the quantity of localised MB was low. The overestimation had a positive gradient during the first minute of US data or at about the first $\sim250$ localised MBs. The maximum overestimation occurred at about $1$ min of US data or $\sim250$ localised MBs. After this, the overestimation decreased as the average of the localised MBs increased. It was observed that the average error and standard deviation plateaus once, on average, $\sim1000$ MBs were localised in the estimated diameter. This suggests that their statistical distribution could be described after $\sim1000$ MBs are localised.    

To plateau the average diameter estimated with SRUS and its standard deviation, an average of $500$ to $1000$ localised MBs per estimated diameter was required. This behaviour may reflect that the average of $500$ to $1000$ localised MBs describe the microvessel's diameter with the SRUS strategy used. Hence, adding extra localised MBs to the estimated diameter would have minimal effect on the statistical distribution.

Improvements in the US data acquisition protocol and SRUS algorithms could reduce the SRUS estimation error and average MB count. For example, adding more angled plane waves could be implemented to reduce the localisation error due to incoherent noise, or a localisation algorithm with greater confidence in its MB signals detection could reduce the required average MB count to plateau.  

\section{Conclusion}
\label{sec:conclusion}

This work presented a methodology for fabricating repeatable, stable, durable, optically validated, and physiologically relevant microvascular phantoms. The methodology allows the fabrication of microvascular phantoms with variable vascular density, vessel diameter, and branches. It was shown that the fabricated microvascular phantoms offer a platform to support detailed investigation and improvements of SRUS methods, i.e., localisation, and microvascular density algorithms, within controlled environments. The phantoms presented offer physiologically relevant microvasculature features with ground truth for SRUS validation. In addition, due to its durability, the phantom allows repeated performance testing and comparison on the same microvascular structure of methodological developments.

The microvascular phantoms comprised a supporting matrix material and a vessel that could bifurcate or branch into several vessels of different diameters. The methodology for manufacturing optically validated microvascular phantoms comprised $6$ steps: microrod extrusion, microrods connection, construction of the microvasculature negative print, moulding the phantom, dissolution of the microvasculature negative print, and MB flow test. The repeatability of the methodology was demonstrated by fabricating bifurcating and branching microvascular phantoms. It was demonstrated that the phantoms are durable; they sustained compression and flow cyclic load testing, and their acoustic and optical properties varied less than $10\%$ after a year of fabrication. The proposed methodology is inexpensive, easy to implement, and can be carried out with common laboratory equipment, such as a 3D printer, a vacuum pump, and a US cleaner. 

A method to evaluate the performance of SRUS as a function of the localised MBs was presented and experimentally tested. The method was used to evaluate the performance of the SRUS strategy used in this work. It was shown that the SRUS performance improved as the number of localised MBs increased, and a minimum of $1000$ localisations per estimated diameter was required for the SRUS performance to plateau. Passing the $1000$ localisations per estimated diameter, an average SRUS error of $7.93$ $\mu$m ($\sim\lambda/44$) with a standard deviation of $11.52$ $\mu$m was achieved.

\section*{Acknowledgment}
The authors would like to thank the members of the QUIIN lab at King’s College London for the helpful feedback and discussions.
\bibliographystyle{IEEetran} 

\bibliography{Bibliography.bib} 

\end{document}